\documentclass[11pt,a4paper]{article}

\usepackage[utf8]{inputenc}
\usepackage[T1]{fontenc}
\usepackage[english]{babel}
\usepackage[a4paper,margin=2.25cm]{geometry}
\usepackage{lmodern}
\usepackage{microtype}
\usepackage{amsmath,amssymb,bm,mathtools}
\usepackage{graphicx}
\usepackage{booktabs,longtable,array,tabularx}
\usepackage{ragged2e}
\usepackage{caption}
\usepackage{pdflscape}
\usepackage{enumitem}
\usepackage[numbers,sort&compress]{natbib}
\makeatletter
\AtBeginDocument{\def\bibitemNoStop{\unskip.\spacefactor\@mmm\space}}
\makeatother
\usepackage{xcolor}
\usepackage[colorlinks=true,linkcolor=blue,citecolor=blue,urlcolor=blue]{hyperref}
\usepackage[nameinlink,noabbrev]{cleveref}
\hypersetup{
  pdftitle={Multipolar static tidal response of Schwarzschild black holes in cubic gravity: a metric-action derivation of tidal running},
  pdfauthor={Edilberto O. Silva},
  pdfsubject={Static black-hole tidal response in cubic gravity},
  pdfkeywords={black holes, tidal Love numbers, effective field theory, higher-curvature gravity, Zerilli--Moncrief equation}
}
\allowdisplaybreaks[3]
\renewcommand{\arraystretch}{1.15}
\newcolumntype{Y}{>{\RaggedRight\arraybackslash}X}
\newcolumntype{C}[1]{>{\Centering\arraybackslash}p{#1}}
\newcolumntype{L}[1]{>{\RaggedRight\arraybackslash}p{#1}}
\newcolumntype{M}[1]{>{\centering\arraybackslash}m{#1}}
\newcolumntype{V}[1]{>{\RaggedRight\arraybackslash}m{#1}}
\newcommand{\actiontablesection}[1]{%
  \addlinespace[0.45em]
  \multicolumn{2}{c}{\itshape #1}\\
  \addlinespace[0.20em]
}

\newcommand{\dd}{\mathrm d}
\newcommand{\rs}{r_s}
\newcommand{\epse}{\epsilon_{\rm e}}
\newcommand{\Mpl}{M_{\rm Pl}}
\newcommand{\Mzero}{M_0}
\newcommand{\Oe}{\mathcal O_{\rm e}}

\newcommand{\betahat}{\widehat\beta}

\newcommand{\order}{\mathcal O}
\newcommand{\affiliation}[1]{\gdef\theaffiliation{#1}}
\newcommand{\theaffiliation}{}
\newcommand{\keywords}[1]{\par\vspace{0.6em}\noindent\textbf{Keywords:} #1\par}

\title{\textbf{Multipolar static tidal response of Schwarzschild black holes in cubic gravity:\\[0.35em]a metric-action derivation of tidal running}}
\affiliation{Programa de P\'os-Gradua\c c\~ao em F\'{\i}sica \& Coordena\c c\~ao do Curso de F\'{\i}sica -- Bacharelado, Universidade Federal do Maranh\~{a}o, 65085-580 S\~{a}o Lu\'{\i}s, Maranh\~{a}o, Brazil}
\author{Edilberto O. Silva\\[2mm]
\small\begin{minipage}{0.92\textwidth}\centering
\theaffiliation
\end{minipage}\\[1mm]
\small \texttt{edilberto.silva@ufma.br}}
\date{}

\begin{document}
\maketitle

\begin{abstract}
Static Love numbers of four-dimensional Schwarzschild black holes vanish in general relativity, whereas higher-curvature operators generate a nontrivial response. We study the parity-even cubic Weyl interaction directly in metric variables and carry out the electric, static calculation for every integer multipole $\ell\geq2$. The angular reduction is organized by $L=\ell(\ell+1)$ and yields radial actions whose coefficients are at most linear in $L$ in the Einstein sectors and quadratic in $L$ in the cubic sector. After perturbative order reduction, the three metric equations become a constrained two-dimensional first-order system. Eliminating one field produces a scalar equation with the general-relativistic static tidal operator and homogeneous basis $P_\ell^{2}(2r/\rs-1)$, $Q_\ell^{2}(2r/\rs-1)$. A Frobenius calculation gives the gauge-invariant Zerilli--Moncrief running coefficient
$\beta_\ell^{\rm ZM}=\epse\,7L^2(L-2)^2(L-4)(L-6)/12$.
The factor $L-6$ makes the quadrupole the only physical electric multipole without logarithmic running. The exact quadrupolar solution yields the fixed-integer Zerilli--Moncrief branch ratio $-2400\epse$, distinct from the analytically continued canonical Love number $k_2^E=448\epse$. For the octupole we construct the global horizon-regular metric solution and show explicitly how logarithms in the two Green-function channels cancel at the horizon while surviving in the asymptotic response. We then derive the normalization map from the Zerilli--Moncrief logarithm to the canonical electric beta function and obtain the corresponding length-scale running of finite-size worldline coefficients. The canonical beta functions agree with the modified-Teukolsky result in the literature. The metric-action route additionally provides the radial action, the full metric reconstruction, and a direct explanation of the exceptional quadrupole.
\end{abstract}


\tableofcontents

\section{Introduction}
\label{sec:introduction}

The response of a self-gravitating body to an external tidal field is one of the cleanest ways of connecting strong-field geometry with the long-distance dynamics of compact binaries. In a worldline effective field theory (EFT), the conservative response is encoded in Wilson coefficients multiplying local tidal operators; in relativistic perturbation theory, the same information is read from the relative amplitudes of the asymptotic source and response branches. These coefficients enter the post-Newtonian and effective-one-body descriptions of inspiral and therefore provide a systematic bridge between the internal structure of compact objects and gravitational-wave observables~\cite{Love1909,GoldbergerRothstein2006,Porto2016,FlanaganHinderer2008,Hinderer2008,YagiYunes2013,CardosoFranzinMaselliPaniRaposo2017,RodriguezSantoniSolomon2026}. The static coefficients are also useful theoretically: they test how horizon regularity, asymptotic matching, gauge invariance, and EFT renormalization cooperate in a genuinely strong-field problem.

Four-dimensional asymptotically flat black holes in general relativity (GR) are exceptional because their static Love numbers vanish. This result was first established by direct metric calculations for Schwarzschild black holes and has since been reformulated in gauge-invariant, EFT, and curvature-based languages~\cite{BinningtonPoisson2009,DamourNagar2009,KolSmolkin2012,HuiJoycePencoSantoniSolomon2021,Chia2021,LeTiecCasalsFranzin2021}. The cancellation is not merely a low-multipole accident. Analytic continuation in the angular momentum, worldline matching, ladder symmetries, and hidden symmetries of the static perturbation equations provide complementary explanations~\cite{HuiJoycePencoSantoniSolomon2022,CharalambousDubovskyIvanov2021,BenAchourLivineMukohyamaUzan2022,RivaSantoniSavicVernizzi2024}. Recent nonlinear analyses have strengthened this picture by showing that broad classes of quadratic and fully nonlinear static responses also vanish in Schwarzschild GR~\cite{IteanuRivaSantoniSavicVernizzi2024,CombaluzierHuiSantoniSolomonWong2024}. The vanishing Love numbers therefore represent a structural property of the theory rather than a generic feature of compact objects.

This special GR result makes tidal response a sensitive diagnostic of new short-distance physics. Higher-curvature operators arise when heavy degrees of freedom are integrated out and provide a controlled derivative expansion of the gravitational action. They modify both the black-hole background and the perturbation equations, and they can generate finite Love numbers or logarithmically running response coefficients even when the GR value is zero~\cite{EndlichGorbenkoHuangSenatore2017,CardosoKimuraMaselliSenatore2018,SennettBritoBuonannoGorbenkoSenatore2020,CanoGanchevMayersonRuiperez2022}. The resulting response is conceptually richer than a single number. At fixed integer multipole, local terms can mix with the decaying homogeneous branch; analytic continuation in $\ell$ provides a canonical source--response split; and logarithmic terms define renormalization-group beta functions whose coefficients are insensitive to finite subtraction conventions.

Several recent approaches have clarified these issues. Gauge-invariant Regge--Wheeler--Zerilli calculations, Green-function methods, EFTs of black-hole perturbations, and modified Teukolsky equations have been used to extract static responses in GR and beyond~\cite{KatagiriIkedaCardoso2024,GharibBaruraEtAl2024,BarbosaBraxFichetDeSouza2025,Cano2025}. In particular, Ref.~\cite{Cano2025} obtained the complete electric, magnetic, and parity-mixing Love numbers and beta functions of nonrotating black holes in a general higher-derivative gravitational EFT, using analytic continuation whenever the finite part at integer $\ell$ is ambiguous. The matching of higher-curvature Love numbers to worldline Wilson coefficients has subsequently been revisited in detail~\cite{WangLehnerMicolSturani2026}. Related work on dynamical tides emphasizes that a consistent connection between black-hole perturbation theory and the worldline EFT remains essential once frequency and spin are included~\cite{ChakrabortySakethHindererSteinhoff2026}.

The canonical beta functions for cubic gravity are already known from the modified-Teukolsky analysis of Ref.~\cite{Cano2025}. Here we pursue a complementary question: what does a derivation from the metric action reveal when the angular eigenvalue
\begin{equation}
 L\equiv \ell(\ell+1)
 \label{eq:Ldefinition}
\end{equation}
is retained throughout the reduction? This route reconstructs the metric perturbation and exposes structures that are compressed in a final master-curvature response formula. We show that the three static even-parity metric equations reduce, in closed form for the integer multipole tower, to a constrained two-dimensional first-order system; that eliminating one field produces exactly the GR static tidal operator; and that the higher-curvature source develops a resonance proportional to $L-6$. The last factor explains directly at the level of the metric source why the quadrupole is the unique physical electric multipole without logarithmic running.

The derivation proceeds in three stages. We first reduce the action and reconstruct its low-degree dependence on $L$ from exact integer-multipole projections. Multipoles not used in the reconstruction, together with symbolic Euler--Lagrange identities, test the resulting expressions. We then order-reduce the metric equations and cast them as a scalar Green-function problem. At that point a Frobenius recurrence yields the logarithmic coefficient algebraically in $L$; the final beta function is not obtained by interpolation. Finally, we express the result in the Zerilli--Moncrief variable, derive its conversion to the canonical modified-Teukolsky normalization, and translate the canonical response into finite-size worldline coefficients.

A central conceptual point is that three related but inequivalent quantities must be kept separate. They are summarized in \cref{tab:responseDefinitions}.
\begin{table}[t]
\centering
\caption{Response quantities used throughout the paper. The distinction between the first two entries is essential for the non-running quadrupole.}
\label{tab:responseDefinitions}
\begin{tabularx}{\textwidth}{@{}L{0.20\textwidth}Y Y@{}}
\toprule
\textbf{Quantity} & \textbf{Definition} & \textbf{Physical role} \\
\midrule
$\Delta(B/A)_\ell^{\rm ZM,fixed}$ & Ratio of the decaying and growing homogeneous Zerilli--Moncrief branches after specializing to integer $\ell$ & Gauge invariant, but its finite part can depend on the fixed-integer subtraction prescription \\
\addlinespace
$k_\ell^E$ & Canonical electric Love number defined by analytic continuation and canonical asymptotic normalization & Appropriate finite response for comparison with the modified-Teukolsky result and standard Love-number conventions \\
\addlinespace
$c_{E,\ell}^{\rm fs}$ & Coefficient of the finite-size electric worldline operator & Exterior finite-size contribution; a complete point-source coefficient may also contain $c_{E,\ell}^{\rm pp}$ \\
\bottomrule
\end{tabularx}
\end{table}
For the quadrupole, the fixed-integer Zerilli--Moncrief ratio $\Delta(B/A)_2^{\rm ZM,fixed}$ and $k_2^E$ are both gauge invariant but numerically different. This is not a contradiction. The canonical finite part is defined by continuing away from integer angular momentum before taking the limit, whereas the fixed-integer branch ratio is extracted after the degeneracy between local terms and the response branch has occurred.

The main results are as follows.
\begin{enumerate}[label=(\roman*),leftmargin=2.2em]
\item We obtain radial actions in closed form for the full integer electric-multipole tower, with the complete coefficient lists supplied in \cref{app:actions}.
\item The order-reduced metric equations form a closed first-order system with no exterior singularity other than the horizon and infinity; the third metric field is reconstructed algebraically.
\item The scalar correction obeys the same homogeneous differential operator as the GR tide. The cubic interaction therefore acts as a resonant source in the $P_\ell^2$--$Q_\ell^2$ basis.
\item The Frobenius recurrence and the Green-function residue yield the factorized running coefficient, showing that $\ell=2$ is the unique physical electric multipole without logarithmic running, whereas every $\ell\geq3$ runs.
\item We construct exact global solutions for $\ell=2$ and $\ell=3$. The octupole gives the first explicit running example and displays a nontrivial cancellation of horizon logarithms between the two Green-function channels.
\item We derive the bridge between the Zerilli--Moncrief and canonical modified-Teukolsky beta functions and obtain the corresponding running of the finite-size multipolar worldline coefficients.
\end{enumerate}

The article is organized as follows. \Cref{sec:setup} defines the EFT, corrected background, perturbations, and GR benchmark. The angular identities and arbitrary-$L$ radial action are developed in \cref{sec:angularAction}. The order-reduced metric equations and scalar master equation are derived in \cref{sec:radialSystem,sec:masterGreen}. The generic running is proved in \cref{sec:runningProof}, and the gauge-invariant extraction is given in \cref{sec:gaugeInvariant}. The exact quadrupole and global octupole solutions are discussed in \cref{sec:quadrupole,sec:octupole}. The bridge to the canonical and finite-size worldline normalizations is established in \cref{sec:canonicalWorldline}. We close with the physical interpretation, limitations, and extensions in \cref{sec:discussion}. Full radial-action coefficients and the longer algebraic formulas are collected in the appendices.

\section{Cubic EFT, corrected background, and GR tidal benchmark}
\label{sec:setup}

We begin by fixing the EFT basis, the background parametrization, and the perturbation conventions. These choices matter for finite response coefficients: a local field redefinition can change the off-shell representative of the cubic interaction, while a redefinition of the mass parameter can reshuffle terms already proportional to the EFT coupling. The logarithmic response is insensitive to such finite choices, but the conventions must be stated before the metric solution is compared with canonical Love numbers.

\subsection{Bulk action and coupling conventions}

We consider the parity-even cubic Weyl theory
\begin{equation}
 S_{\rm bulk}
 =\frac{\Mpl^2}{2}\int \dd^4x\sqrt{-g}
 \left[R+\epse\rs^4\Oe\right],
 \qquad
 \Oe=C_{\mu\nu}{}^{\rho\sigma}
 C_{\rho\sigma}{}^{\alpha\beta}
 C_{\alpha\beta}{}^{\mu\nu},
 \label{eq:bulkAction}
\end{equation}
where $\Mpl^{-2}=8\pi G$ and
\begin{equation}
 \epse\equiv\frac{\lambda_{\rm e}}{\Lambda^4\rs^4},
 \qquad |\epse|\ll1.
 \label{eq:epsilonDefinition}
\end{equation}
The cubic interaction is treated as an EFT correction rather than as an additional fundamental kinetic term. We retain terms through first order in $\epse$ and use the lower-order Einstein equations to remove higher radial derivatives from the corrected equations.

It is useful to distinguish the Schwarzschild mass parameter
\begin{equation}
 \Mzero\equiv\frac{\rs}{2}
 \label{eq:MzeroDefinition}
\end{equation}
from the ADM mass of the corrected solution. For comparison with the Riemann-cubic convention of Ref.~\cite{Cano2025}, we define
\begin{equation}
 \lambda_{\rm ev}=\epse\rs^4,
 \qquad
 \frac{\lambda_{\rm ev}}{\Mzero^4}=16\epse.
 \label{eq:couplingMap}
\end{equation}
The $C^3$ and $R^3$ contractions agree on Ricci-flat configurations. Off shell they differ by Ricci-dependent terms and therefore correspond to different EFT representatives related by local field redefinitions. Equation~\eqref{eq:couplingMap} is used only for the common Ricci-flat representative; finite quantities are interpreted with this qualification understood.

The Schwarzschild value of the invariant fixes the normalization,
\begin{equation}
 \Oe\big|_{\rm Schw}=\frac{12\rs^3}{r^9}
 =\frac{96\Mzero^3}{r^9}.
 \label{eq:cubicInvariantSchw}
\end{equation}

\subsection{Corrected spherical background}

We use the areal-radius gauge
\begin{equation}
 \dd s^2=-A(r)\dd t^2+\frac{\dd r^2}{B(r)}+r^2\dd\Omega^2,
 \label{eq:backgroundMetric}
\end{equation}
with
\begin{align}
 A(r)&=f(r)\left[1+\epse a(r)\right]+\order(\epse^2),
 &
 B(r)&=f(r)\left[1+\epse b(r)\right]+\order(\epse^2),
 \\
 f(r)&=1-\frac{\rs}{r}.
 \label{eq:backgroundExpansion}
\end{align}
Solving the order-reduced spherical equations fixes
\begin{align}
 a(r)&=-2\sum_{n=1}^{5}\left(\frac{\rs}{r}\right)^n
 +4\left(\frac{\rs}{r}\right)^6,
 \\
 b(r)&=-2\sum_{n=1}^{5}\left(\frac{\rs}{r}\right)^n
 +16\left(\frac{\rs}{r}\right)^6.
 \label{eq:backgroundFunctions}
\end{align}
The integration constants have been chosen so that $r$ remains the areal radius, the Killing horizon stays at $r=\rs$, and $t$ is canonically normalized at infinity. In this parametrization the large-$r$ expansion of the lapse is
\begin{equation}
 A(r)=1-\frac{\rs}{r}-2\epse\frac{\rs}{r}
 +6\epse\left(\frac{\rs}{r}\right)^6
 -4\epse\left(\frac{\rs}{r}\right)^7
 +\order(\epse^2),
 \label{eq:backgroundAsymptoticMass}
\end{equation}
so that
\begin{equation}
 Gm_{\rm ADM}=\Mzero(1+2\epse)+\order(\epse^2).
 \label{eq:ADMmassRelation}
\end{equation}
All response quantities derived below are already linear in $\epse$. Replacing $\Mzero$ by $Gm_{\rm ADM}$ inside their overall normalization would therefore change them only at $\order(\epse^2)$, beyond the accuracy of the present calculation.

For reference, the two functions in \eqref{eq:backgroundFunctions} are obtained by inserting \eqref{eq:backgroundExpansion} in the spherical field equations, order-reducing the cubic terms on Schwarzschild, and solving the two independent radial combinations with the boundary conditions just stated. The perturbation calculation uses these functions as fixed background data; no additional spherical integration constant enters the tidal problem.

\subsection{Static even-parity perturbations}

In Regge--Wheeler gauge~\cite{ReggeWheeler1957,Zerilli1970,Moncrief1974,MartelPoisson2005}, a static axisymmetric electric perturbation is written as
\begin{subequations}
\begin{align}
 g_{tt}&=-A(r)\left[1+\eta H_0(r)P_\ell(\cos\theta)\right],
 \\
 g_{rr}&=\frac{1+\eta H_2(r)P_\ell(\cos\theta)}{B(r)},
 \\
 g_{\theta\theta}&=r^2\left[1+\eta K(r)P_\ell(\cos\theta)\right],
 \\
 g_{\phi\phi}&=r^2\sin^2\theta
 \left[1+\eta K(r)P_\ell(\cos\theta)\right].
 \label{eq:metricAnsatz}
\end{align}
\end{subequations}
The bookkeeping parameter $\eta$ counts the tidal amplitude. Rotational invariance ensures that the radial equations depend on the spherical harmonic only through $L=\ell(\ell+1)$; choosing $m=0$ and using the unnormalized Legendre polynomial is therefore a convenient projection convention, not a restriction on the physics.

The radial fields are expanded as
\begin{align}
 H_0&=-H+\epse X_0+\order(\epse^2),
 &
 H_2&=H+\epse X_2+\order(\epse^2),
 \\
 K&=K_{\rm GR}[H]+\epse X_K+\order(\epse^2).
 \label{eq:fieldExpansion}
\end{align}
The sign in $H_0$ follows from factoring $-A$ in $g_{tt}$. The three $X$ fields denote the first-order EFT correction.

\subsection{General-relativistic tidal solution}

At zeroth order the tide is controlled by one radial function,
\begin{equation}
 r(r-\rs)H''+(2r-\rs)H'
 -\left[L+\frac{\rs^2}{r(r-\rs)}\right]H=0,
 \label{eq:GRtidalEquation}
\end{equation}
with
\begin{equation}
 K_{\rm GR}[H]
 =\frac{\left[(L-2)r^2-(L-4)r\rs-\rs^2\right]H
 +\rs r(r-\rs)H'}{(L-2)r(r-\rs)}.
 \label{eq:KGRconstraint}
\end{equation}
Introducing
\begin{equation}
 x\equiv\frac{r}{\rs},
 \qquad z\equiv2x-1,
 \label{eq:dimensionlessVariables}
\end{equation}
turns \eqref{eq:GRtidalEquation} into the associated-Legendre equation of order two. We use the convention
\begin{equation}
 P_2^2(z)=3(1-z^2),
 \label{eq:LegendreConventionMain}
\end{equation}
with the real exterior branch of $Q_\ell^2(z)$ for $z>1$. The horizon-regular black-hole tide is
\begin{equation}
 H_{P,\ell}(x)\equiv P_\ell^2(2x-1),
 \label{eq:regularGRTide}
\end{equation}
whereas
\begin{equation}
 H_{Q,\ell}(x)\equiv Q_\ell^2(2x-1)
 \label{eq:decayingGRTide}
\end{equation}
is the independent asymptotically decaying branch. For integer $\ell$, $H_{P,\ell}$ is a polynomial times $x(x-1)$ and is regular at the future horizon, while $H_{Q,\ell}$ diverges there. Unless stated otherwise, $H$ in the sourced equations below denotes $H_{P,\ell}$.

\section{Angular reduction and the arbitrary-multipole radial action}
\label{sec:angularAction}

The metric calculation becomes manageable once the angular dependence is separated before the radial equations are solved. We retain $L=\ell(\ell+1)$ throughout this reduction, with the understanding that the physical tower consists of integer $\ell\geq2$. The result is a single closed expression for that tower rather than a collection of unrelated fixed-multipole equations. The harmonic identities below also explain why the dependence on $L$ is polynomial and of low degree.

\subsection{Harmonic identities and polynomial dependence on \texorpdfstring{$L$}{L}}

Let $Y\equiv P_\ell(\cos\theta)$ and denote by $D_A$ the covariant derivative of the unit two-sphere. The radial reduction can be organized without expanding every Legendre polynomial explicitly because all scalar, vector, and tensor contractions reduce to the eigenvalue $L$. With
\begin{equation}
 \mathcal N_\ell\equiv\int \dd\Omega\,Y^2=\frac{4\pi}{2\ell+1},
 \label{eq:harmonicNorm}
\end{equation}
we use
\begin{subequations}
\begin{align}
 D^2Y&=-LY,
 \\
 \int\dd\Omega\,D_A YD^A Y&=L\mathcal N_\ell,
 \\
 \int\dd\Omega\,(D_A D_B Y)(D^A D^B Y)&=L(L-1)\mathcal N_\ell,
 \\
 \int\dd\Omega\,Y_{AB}Y^{AB}&=\frac12L(L-2)\mathcal N_\ell,
 \label{eq:harmonicIdentities}
\end{align}
\end{subequations}
where
\begin{equation}
 Y_{AB}\equiv D_A D_B Y+\frac{L}{2}\Omega_{AB}Y
 \label{eq:tensorHarmonic}
\end{equation}
is the traceless even-parity tensor harmonic. The third identity in \eqref{eq:harmonicIdentities} follows from integration by parts and the unit-sphere Ricci tensor, $R_{AB}=\Omega_{AB}$; the last then follows by subtracting the trace.

Rotational invariance removes any dependence on $m$ after the angular integral. Integration by parts on the sphere then reduces every contraction to the scalar eigenvalue $L$ or to the tensor combination $L(L-2)$. Before the metric constraints are solved, no inverse angular operator is introduced; the coefficient functions are consequently polynomials, rather than rational functions, of $L$. The quadratic Einstein action contains at most one contracted pair of angular derivatives and is therefore at most linear in $L$. The quadratic expansion of the cubic-curvature term contains at most two such pairs and is at most quadratic in $L$. These degree bounds are what make the reconstruction from a small number of integer multipoles mathematically determined rather than an empirical fit.

For the computer-algebra implementation, the angular integral was also evaluated directly. If $u=\cos\theta$ is introduced after the determinant and angular factors have been expanded, the exact change of variables is
\begin{equation}
 \int_0^\pi\dd\theta\,\mathfrak L(r,\theta)
 =\int_{-1}^{1}\frac{\dd u}{\sqrt{1-u^2}}\,
 \mathfrak L(r,\arccos u).
 \label{eq:angularJacobian}
\end{equation}
The direct-$\theta$ projection and the $u$ projection with \eqref{eq:angularJacobian} agree exactly in the benchmark multipoles.

\subsection{Quadratic radial action}

Expanding \eqref{eq:bulkAction} to second order in the tidal amplitude and first order in $\epse$, integrating over the angles, and dividing by \eqref{eq:harmonicNorm} gives
\begin{equation}
 \mathcal L_2^{\rm radial}(L)
 =\mathcal L_{\rm EH}^{(0)}(L)
 +\epse\left[
 \mathcal L_{\rm EH}^{(1)}(L)
 +\rs^4\mathcal L_{C^3}^{(0)}(L)
 \right].
 \label{eq:radialAction}
\end{equation}
The three terms have distinct origins. The first is the Einstein--Hilbert action on Schwarzschild. The second is the Einstein--Hilbert action evaluated on the $\order(\epse)$-corrected background. The cubic term already carries $\epse$ and is therefore evaluated on the GR background and GR tidal ansatz at this order.

We use the coefficient representation
\begin{equation}
 \mathcal L_s(L)=\sum_{\mathcal M}c_{\mathcal M}^{(s)}(r,\rs;L)\,\mathcal M,
 \label{eq:actionCoefficientRepresentation}
\end{equation}
where $\mathcal M$ runs over monomials in $H_0,H_2,K$ and their first and second radial derivatives. The complete coefficient tables are given in \cref{app:actions}. Their global structure is summarized in \cref{tab:actionStructure}.

\begin{table}[t]
\centering
\caption{Structure and validation of the arbitrary-$L$ radial actions. The degree is the largest polynomial degree in $L$ among all coefficient functions.}
\label{tab:actionStructure}
\begin{tabular}{@{}lcccc@{}}
\toprule
\textbf{Sector} & \textbf{Monomials} & \textbf{Degree 0} & \textbf{Degree 1} & \textbf{Degree 2} \\
\midrule
$\mathcal L_{\rm EH}^{(0)}$ & 22 & 19 & 3 & 0 \\
$\mathcal L_{\rm EH}^{(1)}$ & 23 & 20 & 3 & 0 \\
$\mathcal L_{C^3}^{(0)}$ & 36 & 12 & 18 & 6 \\
\bottomrule
\end{tabular}
\end{table}

A few coefficients illustrate how the angular factors enter:
\begin{equation}
 c_{H_0K}^{\rm EH0}=\frac{L-2}{2},
 \qquad
 c_{H_2K}^{\rm EH0}=\frac{L+2}{2},
 \qquad
 c_{K^2}^{C^3}=\frac{\rs}{r^5}(L-2)^2.
 \label{eq:representativeActionCoefficients}
\end{equation}
The factors $L-2=(\ell-1)(\ell+2)$ reflect the tensorial angular structure of even-parity gravitational perturbations rather than an accidental interpolation pattern.

\subsection{Reconstruction and independent checks}

The degree bounds implied by \eqref{eq:harmonicIdentities} fix a linear coefficient from two integer multipoles and a quadratic coefficient from three. Multipoles not used in that determination were then projected independently. For every $2\leq\ell\leq13$, the directly projected action agrees coefficient by coefficient with the closed expression in $L$.

Two dynamical checks go beyond this coefficient comparison. Inserting the GR equation \eqref{eq:GRtidalEquation} and constraint \eqref{eq:KGRconstraint} into the Euler--Lagrange equations of $\mathcal L_{\rm EH}^{(0)}(L)$ reduces them identically to zero with $L$ symbolic. Likewise, once the EFT equations are written in the form derived in \cref{sec:radialSystem}, substituting that system back into the three original metric equations reduces each equation identically to zero. Thus the reconstructed action is tested at the level of its radial dynamics, not only through the sequences of angular coefficients.

\subsection{Generalized Euler--Lagrange equations and order reduction}

Because the cubic contribution contains second radial derivatives in the reduced density, the appropriate one-dimensional variational operator is
\begin{equation}
 \frac{\delta\mathcal L}{\delta q}
 =\sum_{n=0}^{N}(-1)^n\frac{\dd^n}{\dd r^n}
 \left(\frac{\partial\mathcal L}{\partial q^{(n)}}\right),
 \qquad q\in\{H_0,H_2,K\}.
 \label{eq:generalizedEL}
\end{equation}
At first order in $\epse$, the field equations have the schematic form
\begin{equation}
 \mathcal D_{\rm GR}\bm X
 =-\mathcal E_{\rm EH}^{(1)}[H]
 -\rs^4\mathcal E_{C^3}^{(0)}[H],
 \label{eq:linearizedEFTEquation}
\end{equation}
where $\mathcal D_{\rm GR}$ is the linearized Einstein operator acting on $(X_0,X_2,X_K)$. Any $H^{(n)}$ with $n\geq2$ in the source is reduced recursively with \eqref{eq:GRtidalEquation} and its derivatives. This is the perturbative field-redefinition step that removes higher-derivative modes outside the domain of the truncated EFT. The coefficient tables retain a representative with second radial derivatives, so the variation is understood with compact support in the bulk. Total radial derivatives, together with the boundary terms required for a well-posed gravitational variational problem, do not change the local Euler--Lagrange equations used here.

\section{Order-reduced metric equations}
\label{sec:radialSystem}

The three radial functions in the ansatz are not independent static degrees of freedom. The field equations contain a constraint that can be used to reconstruct one function algebraically, leaving a two-component first-order system. Besides simplifying the boundary-value problem, this form makes the horizon data transparent and prepares the scalar reduction used for the response.

We now set $x=r/\rs$ and use a prime for $\dd/\dd x$.

\subsection{Closed first-order system}

Define
\begin{equation}
 \bm X(x)=\begin{pmatrix}X_0(x)\\X_K(x)\end{pmatrix}.
 \label{eq:Xvector}
\end{equation}
The order-reduced equations are
\begin{equation}
 \bm X'=M_L(x)\bm X+\bm s_H(L,x)H+\bm s_{H'}(L,x)H',
 \label{eq:dimensionlessFirstOrderSystem}
\end{equation}
where
\begin{equation}
 M_L(x)=
 \begin{pmatrix}
 \dfrac{(2-L)x^2+(L-4)x+1}{x(x-1)} & 2-L
 \\[3mm]
 \dfrac{(L-2)x+2}{x} & L-2
 \end{pmatrix}.
 \label{eq:dimensionlessMatrix}
\end{equation}
The source vectors are recorded in \cref{app:sources}. They follow directly from the radial action and are rational in $x$, with numerators no higher than quadratic in $L$.

The remaining metric correction is
\begin{align}
 X_2={}-X_0
 +\frac{12\left[(L+24)x^2-(L+62)x+39\right]}{x^6(x-1)}H
 -\frac{12(14x-15)}{x^5}H'.
 \label{eq:X2AlgebraicDimensionless}
\end{align}
One convenient elimination order is to solve two metric equations for $X_2'$ and $X_0'$, insert those expressions into the $K$ equation, and then solve for $X_K'$. The unused combination contains no derivatives after these substitutions and gives \eqref{eq:X2AlgebraicDimensionless}. Conversely, inserting \eqref{eq:dimensionlessFirstOrderSystem} and \eqref{eq:X2AlgebraicDimensionless} in the three Euler--Lagrange equations reduces each one identically to zero.

\subsection{Horizon regularity and independent data}

The residue matrix at $x=1$ is
\begin{equation}
 \lim_{x\to1}(x-1)M_L(x)
 =\begin{pmatrix}-1&0\\0&0\end{pmatrix}.
 \label{eq:horizonResidueMatrix}
\end{equation}
Thus the homogeneous system has one singular and one regular local exponent. The geometric content of the regularity condition is easiest to see in ingoing coordinates. Define
\begin{equation}
 \dd v=\dd t+\frac{\dd r}{\sqrt{A(r)B(r)}}.
 \label{eq:ingoingCoordinate}
\end{equation}
At linear order in the tidal perturbation, the coefficient of $\dd r^2$ in the $(v,r)$ chart contains
\begin{equation}
 \delta g_{rr}^{(v,r)}
 =\eta\,\frac{H_2-H_0}{B}\,Y+\text{terms finite at }r=\rs.
 \label{eq:ingoingRegularityCombination}
\end{equation}
Since $B\sim r-\rs$, regularity requires $H_2-H_0=\order(r-\rs)$. The GR part obeys this condition because $H=\order(r-\rs)$, and at order $\epse$ it gives
\begin{equation}
 X_2(1)=X_0(1).
 \label{eq:horizonEqualityGeometric}
\end{equation}

For the Legendre normalization,
\begin{equation}
 H_{P,\ell}(x)=h_1(x-1)+\order((x-1)^2),
 \qquad
 h_1=-\frac12L(L-2).
 \label{eq:horizonHcoefficient}
\end{equation}
The leading terms of the sourced equations then fix
\begin{equation}
 X_0(1)=12h_1=-6L(L-2),
 \qquad
 X_2(1)=X_0(1),
 \label{eq:genericHorizonData}
\end{equation}
while $X_K(1)$ is the one regular homogeneous datum. The exact solutions below show that this constant changes the normalization of the regular growing tide; it is not an independent black-hole response parameter. Equation~\eqref{eq:genericHorizonData} gives $-144$ at $\ell=2$ and $-720$ at $\ell=3$.

At infinity,
\begin{equation}
 \lim_{x\to\infty}M_L(x)
 =(L-2)\begin{pmatrix}-1&-1\\1&1\end{pmatrix},
 \qquad
 \left[\lim_{x\to\infty}M_L(x)\right]^2=0.
 \label{eq:nilpotentInfinityMatrix}
\end{equation}
The limiting matrix is nilpotent, so its eigenvalues do not determine the asymptotic powers by themselves; the subleading $1/x$ terms distinguish the growing and decaying branches. The only denominators in the exterior system are powers of $x$ and $x-1$. In particular, no pole occurs at the photon sphere, and no extra regularity condition is introduced there.

\section{Scalar master equation and Green-function representation}
\label{sec:masterGreen}

The first-order system is useful for reconstructing the metric, whereas the source--response split is clearest in a second-order equation. The reduction below shows explicitly that the cubic interaction changes the source but not the homogeneous tidal operator at this order.

\subsection{Elimination of \texorpdfstring{$X_K$}{XK}}

Write
\begin{equation}
 S_i\equiv s_{H,i}H+s_{H',i}H',
 \qquad i=1,2.
 \label{eq:SiDefinition}
\end{equation}
The two rows of \eqref{eq:dimensionlessFirstOrderSystem} are
\begin{align}
 X_0'&=M_{11}X_0+M_{12}X_K+S_1,
 \\
 X_K'&=M_{21}X_0+M_{22}X_K+S_2.
 \label{eq:firstOrderRows}
\end{align}
Since $M_{12}=2-L$ is constant and nonzero for the physical multipoles, the first row gives
\begin{equation}
 X_K=\frac{X_0'-M_{11}X_0-S_1}{M_{12}}
 =-\frac{X_0'-M_{11}X_0-S_1}{L-2}.
 \label{eq:XKfromX0}
\end{equation}
Differentiating and using the second row yields the identity
\begin{align}
 X_0''={}&(M_{11}+M_{22})X_0'
 +\left(M_{11}'+M_{12}M_{21}-M_{22}M_{11}\right)X_0
 \notag\\
 &+S_1'+M_{12}S_2-M_{22}S_1.
 \label{eq:scalarEliminationIdentity}
\end{align}
For the matrix \eqref{eq:dimensionlessMatrix},
\begin{align}
 -(M_{11}+M_{22})&=\frac{2x-1}{x(x-1)},
 \\
 -\left(M_{11}'+M_{12}M_{21}-M_{22}M_{11}\right)
 &=-\frac{Lx(x-1)+1}{x^2(x-1)^2}.
 \label{eq:scalarOperatorCoefficients}
\end{align}
Reducing $H''$ in the source term of \eqref{eq:scalarEliminationIdentity} with the GR equation therefore gives
\begin{equation}
 X_0''+\frac{2x-1}{x(x-1)}X_0'
 -\frac{Lx(x-1)+1}{x^2(x-1)^2}X_0
 =j_L[H,H'].
 \label{eq:scalarMasterDimensionless}
\end{equation}
The source has the form
\begin{equation}
 j_L[H,H']=j_H(L,x)H+j_{H'}(L,x)H',
 \label{eq:sourceDecomposition}
\end{equation}
with $j_H$ and $j_{H'}$ given in \cref{eq:explicitjH,eq:explicitjHp}.

The homogeneous part of \eqref{eq:scalarMasterDimensionless} is the GR operator in \eqref{eq:GRtidalEquation}. In terms of $z=2x-1$,
\begin{equation}
 (z^2-1)X_{0,zz}+2zX_{0,z}
 -\left[L+\frac{4}{z^2-1}\right]X_0=0,
 \label{eq:associatedLegendreEquation}
\end{equation}
so a fundamental basis is
\begin{equation}
 u_\ell(x)=P_\ell^2(2x-1),
 \qquad
 v_\ell(x)=Q_\ell^2(2x-1),
 \label{eq:homogeneousBasis}
\end{equation}
with Wronskian
\begin{equation}
 W_x[u_\ell,v_\ell]
 =\frac{L(L-2)}{2x(x-1)}.
 \label{eq:Wronskian}
\end{equation}

\subsection{Variation of parameters}

A particular solution is
\begin{align}
 X_0^{\rm part}(x)={}&-u_\ell(x)
 \int^x\frac{v_\ell(t)j_L(t)}{W_t}\,\dd t
 \notag\\
 &+v_\ell(x)
 \int^x\frac{u_\ell(t)j_L(t)}{W_t}\,\dd t.
 \label{eq:variationParameters}
\end{align}
The lower limits are chosen jointly so that the full expression, rather than each channel separately, is regular at the horizon. A remaining multiple of $u_\ell$ is fixed by demanding that the EFT correction not change the applied tidal amplitude.

The coefficient of the decaying branch is governed by
\begin{equation}
 I_{Q,\ell}'(x)
 =\frac{u_\ell(x)j_L(x)}{W_x}
 =\frac{2x(x-1)}{L(L-2)}H_{P,\ell}(x)j_L(x).
 \label{eq:QchannelIntegrand}
\end{equation}
For integer $\ell$, $H_{P,\ell}$ is polynomial. Once the GR recurrence is imposed, $I_{Q,\ell}'$ is a finite Laurent-rational function with possible poles only at $x=0$ and $x=1$,
\begin{equation}
 I_{Q,\ell}'(x)
 =\sum_{n=-N}^{N'}c_nx^n
 +\sum_{m=1}^{M}\frac{d_m}{(x-1)^m}.
 \label{eq:LaurentDecompositionIQ}
\end{equation}
Integration produces $c_{-1}\log x+d_1\log(x-1)$. The horizon calculation in \cref{sec:horizonNoRunning} gives $d_1=0$; the surviving asymptotic logarithm is therefore determined by the residue $c_{-1}$ at the algebraic singular point $x=0$. No physical boundary condition is imposed at $x=0$, which lies inside the black hole. It is used only as an efficient way to read the global Laurent coefficient that integrates to $\log x$.

The other channel,
\begin{equation}
 I_{P,\ell}'(x)=-\frac{v_\ell(x)j_L(x)}{W_x},
 \label{eq:PchannelIntegrand}
\end{equation}
controls the growing branch. The separate products in \eqref{eq:variationParameters} may contain horizon logarithms even though their sum is analytic; the octupolar solution in \cref{sec:octupole} displays this cancellation explicitly.

\section{Algebraic proof of multipolar tidal running}
\label{sec:runningProof}

The full global solution is not needed to determine the logarithm. For integer $\ell$, the regular GR tide is polynomial and the Green-function integrand has the finite Laurent form described in \eqref{eq:LaurentDecompositionIQ}. Its $x^{-1}$ coefficient fixes the asymptotic logarithm, while a separate expansion at the horizon tests whether a second logarithm is generated there. The calculation can therefore be performed algebraically in $L$, after the closed action for the integer tower has been obtained.

\subsection{Frobenius recurrence at \texorpdfstring{$x=0$}{x=0}}

Write the regular homogeneous solution as
\begin{equation}
 H(x)=\sum_{n=1}^{\infty}a_nx^n.
 \label{eq:FrobeniusSeries}
\end{equation}
Substitution into the dimensionless GR equation gives the recurrence
\begin{align}
 (n^2-1)a_n
 +\left[L-(n-1)(2n-1)\right]a_{n-1}
 +\left[(n-2)(n-1)-L\right]a_{n-2}=0,
 \qquad n\geq2,
 \label{eq:FrobeniusRecurrence}
\end{align}
with $a_0=0$ and $a_1$ free. The first coefficients are
\begin{subequations}
\begin{align}
 a_2&=-\frac{L-3}{3}a_1,
 \\
 a_3&=\frac{(L-6)(L-4)}{24}a_1,
 \\
 a_4&=-\frac{(L-12)(L-6)(L-5)}{360}a_1,
 \\
 a_5&=\frac{(L-20)(L-12)(L-6)^2}{8640}a_1,
 \\
 a_6&=-\frac{(L-30)(L-20)(L-12)(L-7)(L-6)}{302400}a_1.
 \label{eq:FrobeniusCoefficients}
\end{align}
\end{subequations}
The repeated appearance of $L-6$ already signals a special truncation at the quadrupole.

For the Legendre normalization \eqref{eq:regularGRTide}, the leading coefficient follows directly from
\begin{equation}
 P_\ell^2(z)=(1-z^2)\frac{\dd^2P_\ell(z)}{\dd z^2}
 \label{eq:LegendreDerivativeIdentity}
\end{equation}
and
\begin{equation}
 P_\ell''(-1)=(-1)^\ell\frac{(\ell+2)!}{8(\ell-2)!}.
 \label{eq:LegendreEndpointDerivative}
\end{equation}
Since $z=2x-1$ and $1-z^2=4x(1-x)$, one obtains
\begin{equation}
 a_1=(-1)^\ell\frac{(\ell-1)\ell(\ell+1)(\ell+2)}{2}
 =(-1)^\ell\frac{L(L-2)}{2}.
 \label{eq:a1Normalization}
\end{equation}

\subsection{Residue of the decaying Green-function channel}

Using the explicit source in \cref{app:sources}, the coefficient of $x^{-1}$ in the numerator $x(x-1)Hj_L$ can first be written as a polynomial in $a_1,\ldots,a_5$. The complete intermediate expression is displayed in \cref{eq:rawResiduePolynomial}. Substituting the recurrence coefficients \eqref{eq:FrobeniusCoefficients} collapses that polynomial to
\begin{equation}
 \mathcal R_L(a_1)
 =-\frac{7}{6}a_1^2L(L-2)(L-4)(L-6).
 \label{eq:residueA1}
\end{equation}
No interpolation of the response is involved in this step. The identity is algebraic in $L$ and therefore applies to the full integer multipole tower represented by $L=\ell(\ell+1)$.

After inserting \eqref{eq:a1Normalization},
\begin{equation}
 \mathcal R_L
 =-\frac{7}{24}L^3(L-2)^3(L-4)(L-6).
 \label{eq:closedResidue}
\end{equation}
The Wronskian factor in \eqref{eq:QchannelIntegrand} converts the residue into the logarithmic coefficient of the scalar metric field,
\begin{equation}
 \betahat_{X_0}(L)
 =-\frac{7}{12}L^2(L-2)^2(L-4)(L-6),
 \label{eq:betaX0}
\end{equation}
where $\beta_{X_0}=\epse\betahat_{X_0}$. Reconstructing the spatial fields with \eqref{eq:dimensionlessFirstOrderSystem} and \eqref{eq:X2AlgebraicDimensionless} gives
\begin{equation}
 \betahat_{X_K}(L)=\betahat_{X_2}(L)
 =-\betahat_{X_0}(L).
 \label{eq:betaSpatial}
\end{equation}

\subsection{Why the horizon does not generate the running}
\label{sec:horizonNoRunning}

Set $x=1+y$ and write $H=b_1y+b_2y^2+b_3y^3+\cdots$. The GR equation gives $b_2=(L-3)b_1/3$ and $b_3=(L^2-10L+24)b_1/24$. Substituting these coefficients in $I_{Q,\ell}'$ removes the $y^{-1}$ term. In the notation of \eqref{eq:LaurentDecompositionIQ}, $d_1=0$. The poles of order two and higher determine regular particular data after the two Green-function channels are combined, but they do not integrate to a horizon logarithm. The scale dependence is therefore an asymptotic response effect, not a failure of future-horizon regularity.

The physical multipoles satisfy $L\geq6$. Equation~\eqref{eq:betaX0} therefore implies
\begin{equation}
 \ell=2:\quad \betahat_{X_0}=\betahat_{X_K}=\betahat_{X_2}=0,
 \qquad
 \ell\geq3:\quad \betahat_{X_K}=\betahat_{X_2}>0
 \label{eq:quadrupoleException}
\end{equation}
for positive $\epse$. The quadrupole is the unique physical electric multipole for which the cubic source is nonresonant. This conclusion is stronger than observing a zero in a list of beta functions: the factor $L-6$ has been traced to the local recurrence of the GR tide and the explicit metric source.

\section{Gauge-invariant Zerilli--Moncrief response}
\label{sec:gaugeInvariant}

The reduced variables are defined in Regge--Wheeler gauge, whereas the response coefficient should not depend on that choice. We therefore reconstruct the Zerilli--Moncrief invariant and follow the logarithmic decaying branch through the metric-to-master-variable map. This also fixes the normalization needed in \cref{sec:canonicalWorldline}.

\subsection{Covariant even-parity invariants}

Let $x^a=(t,r)$ denote coordinates on the two-dimensional orbit space, let $Y=Y_{\ell m}$, and define
\begin{equation}
 Y_A\equiv D_A Y,
 \qquad
 Y_{AB}\equiv D_A D_B Y+\frac{L}{2}\Omega_{AB}Y,
 \qquad
 r_a\equiv\nabla_a r.
 \label{eq:gaugeHarmonicDefinitions}
\end{equation}
The even-parity perturbation is decomposed as
\begin{align}
 h_{ab}&=h_{ab}^{\ell m}Y,
 &
 h_{aB}&=j_a^{\ell m}Y_B,
 \\
 h_{AB}&=r^2\left(K^{\ell m}\Omega_{AB}Y
 +G^{\ell m}Y_{AB}\right).
 \label{eq:generalEvenParityDecomposition}
\end{align}
Following Moncrief and Martel--Poisson~\cite{Moncrief1974,MartelPoisson2005}, set
\begin{equation}
 p_a=j_a-\frac12r^2\nabla_aG,
 \label{eq:paDefinition}
\end{equation}
and
\begin{equation}
 \widetilde h_{ab}=h_{ab}-\nabla_ap_b-\nabla_bp_a,
 \qquad
 \widetilde K=K+\frac{L}{2}G-\frac{2}{r}r^ap_a.
 \label{eq:MPinvariants}
\end{equation}
The shift of $p_a$ under an even-parity gauge transformation cancels the corresponding shifts of $h_{ab}$ and $K$. In Regge--Wheeler gauge, $j_a=G=0$, and the invariant fields coincide with the metric functions used in the radial action.

The Martel--Poisson master variable is
\begin{equation}
 \Psi_\ell^{\rm ZM}
 =\frac{2r}{L}\left[
 \widetilde K+\frac{2}{\Lambda_\ell}
 \left(r^ar^b\widetilde h_{ab}-r r^a\nabla_a\widetilde K\right)
 \right],
 \qquad
 \Lambda_\ell=L-2+\frac{3\rs}{r}.
 \label{eq:covariantZM}
\end{equation}
For a static perturbation in Regge--Wheeler gauge this becomes
\begin{equation}
 \Psi_\ell^{\rm ZM}
 =\frac{2r}{L}\left[
 K+\frac{2f}{\Lambda_\ell}(H_2-rK_{,r})
 \right].
 \label{eq:RWZM}
\end{equation}
The covariant variable has dimensions of length. Whenever an expression is written only in terms of $x=r/\rs$ below, we use the dimensionless quantity $\Psi_\ell^{\rm ZM}/\rs$ and suppress the explicit factor of $\rs$. Appendix~\ref{app:gauge} verifies the construction with an arbitrary pure-gauge perturbation.

\subsection{Background dressing at order \texorpdfstring{$\epse$}{epsilon}}

The order-$\epse\eta$ radial perturbation entering \eqref{eq:RWZM} is not $X_2$ alone. Expanding
\begin{equation}
 g_{rr}=\frac{1+\eta(H+\epse X_2)Y}{f(1+\epse b)}
 \label{eq:grrExpansionStart}
\end{equation}
gives
\begin{equation}
 g_{rr}=\frac{1}{f}\left[1-\epse b+\eta HY
 +\epse\eta(X_2-bH)Y\right]+\order(\epse^2,\eta^2).
 \label{eq:grrExpanded}
\end{equation}
Hence
\begin{equation}
 H_{2,\rm phys}^{(1)}=X_2-bH.
 \label{eq:backgroundDressing}
\end{equation}
The subtraction is fixed by the corrected background and is separate from a linear gauge transformation. It is numerically important in the quadrupolar benchmark.

\subsection{Inheritance of the logarithmic coefficient}

The logarithm multiplies the decaying GR branch, not the regular tidal branch. Define
\begin{equation}
 K_{Q,\ell}\equiv K_{\rm GR}[H_{Q,\ell}],
 \qquad
 \Psi_{Q,\ell}^{\rm ZM}\equiv
 \Psi^{\rm ZM}[H_{Q,\ell},K_{Q,\ell}].
 \label{eq:QbranchDefinitions}
\end{equation}
The logarithmic spatial fields obtained from \eqref{eq:betaSpatial} have the form
\begin{equation}
 X_2^{\log}=\betahat_\ell H_{Q,\ell}\log x,
 \qquad
 X_K^{\log}=\betahat_\ell K_{Q,\ell}\log x,
 \qquad
 \betahat_\ell\equiv\betahat_{X_K}=\betahat_{X_2}.
 \label{eq:spatialLogShape}
\end{equation}
Inserting \eqref{eq:spatialLogShape} in the linear map \eqref{eq:RWZM} gives
\begin{equation}
 \delta\Psi_\ell^{\log}
 =\betahat_\ell\Psi_{Q,\ell}^{\rm ZM}\log x
 +\delta\Psi_\ell^{\rm local},
 \label{eq:logInheritance}
\end{equation}
where differentiating the logarithm produces only the nonlogarithmic term
\begin{equation}
 \delta\Psi_\ell^{\rm local}
 =-\frac{4\betahat_\ell x(x-1)K_{Q,\ell}(x)}
 {L[3+(L-2)x]}.
 \label{eq:localZMterm}
\end{equation}
Thus the coefficient of the decaying logarithmic branch is unchanged by the derivative in the Zerilli--Moncrief map. Combining this observation with \eqref{eq:betaSpatial} gives
\begin{equation}
 \beta_\ell^{\rm ZM}=\epse\betahat_\ell,
 \qquad
 \betahat_\ell=\frac{7}{12}L^2(L-2)^2(L-4)(L-6).
 \label{eq:betaZM}
\end{equation}
The scalar coefficient $\betahat_{X_0}$ has the opposite sign; the definition in \eqref{eq:betaZM} follows the common spatial and Zerilli--Moncrief branch.

The decaying master solution is normalized by
\begin{equation}
 \Psi_{Q,\ell}^{\rm ZM}
 \sim\mathcal N_\ell x^{-\ell},
 \qquad
 \mathcal N_\ell
 =\frac{(\ell+1)(\ell+2)\Gamma(\ell+1)^2}
 {(\ell-1)\ell\,\Gamma(2\ell+2)}.
 \label{eq:ZMdecayingNormalization}
\end{equation}
Let $r_0$ be a subtraction length and $x_0=r_0/\rs$. We write the response as
\begin{equation}
 \left[\beta_\ell^{\rm ZM}\log\!\left(\frac{x}{x_0}\right)
 +B_\ell^{\rm ZM}(x_0)\right]\Psi_{Q,\ell}^{\rm ZM}.
 \label{eq:ZMresponseParametrization}
\end{equation}
Independence of $r_0$ gives the length-scale equation
\begin{equation}
 x_0\frac{\dd B_\ell^{\rm ZM}}{\dd x_0}
 =\beta_\ell^{\rm ZM}.
 \label{eq:ZMRGlaw}
\end{equation}
If instead one uses the energy scale $\mu_{\rm R}=r_0^{-1}$, then
\begin{equation}
 \mu_{\rm R}\frac{\dd B_\ell^{\rm ZM}}{\dd\mu_{\rm R}}
 =-\beta_\ell^{\rm ZM}.
 \label{eq:ZMRGenergyLaw}
\end{equation}
This distinction fixes the sign convention used later for the worldline coefficient. The finite $B_\ell^{\rm ZM}$ depends on the subtraction prescription; $\beta_\ell^{\rm ZM}$ does not.

The first values are
\begin{table}[t]
\centering
\caption{Zerilli--Moncrief logarithmic coefficient and decaying-branch normalization for the first electric multipoles.}
\label{tab:firstZMbetas}
\begin{tabular}{@{}cccc@{}}
\toprule
$\ell$ & $L$ & $\betahat_\ell$ & $\mathcal N_\ell$ \\
\midrule
2 & 6  & $0$ & $1/5$ \\
3 & 12 & $403{,}200$ & $1/42$ \\
4 & 20 & $16{,}934{,}400$ & $1/252$ \\
5 & 30 & $256{,}838{,}400$ & $1/1320$ \\
6 & 42 & $2{,}252{,}275{,}200$ & $1/6435$ \\
\bottomrule
\end{tabular}
\end{table}
The rapid growth in the third column belongs to the unnormalized metric branch; the canonical conversion in \cref{sec:canonicalWorldline} carries compensating factorial factors.

\section{Exact quadrupole and the fixed-integer response}
\label{sec:quadrupole}

At $L=6$ the logarithmic coefficient vanishes, so the response is entirely finite and the distinction between prescriptions cannot be hidden in a running term. The quadrupole is therefore both a check of the generic equations and the cleanest place to separate the fixed-integer Zerilli--Moncrief ratio from the canonically continued Love number.

For $\ell=2$, $L=6$, the GR radial functions are
\begin{equation}
 H_2^{\rm GR}=P_2^2(2x-1)=-12x(x-1),
 \qquad
 K^{\rm GR}=K_{\rm GR}[H_2^{\rm GR}]=-6(2x^2-1).
 \label{eq:quadrupoleGR}
\end{equation}
The exact horizon-regular family is
\begin{subequations}
\begin{align}
 X_0={}&-(24+\alpha_1)x^2+\alpha_1x
 +\frac{192}{x^3}+\frac{144}{x^4}-\frac{456}{x^5},
 \\
 X_K={}&(24+\alpha_1)x^2+12-\frac{\alpha_1}{2}
 -\frac{480}{x^3}-\frac{492}{x^4}+\frac{864}{x^5},
 \\
 X_2={}&(24+\alpha_1)x^2-\alpha_1x
 -\frac{480}{x^3}+\frac{3312}{x^4}-\frac{3000}{x^5}.
 \label{eq:quadrupoleFamily}
\end{align}
\end{subequations}
The remaining regular horizon datum is $q_0=X_K(1)$, with
\begin{equation}
 \alpha_1=144+2q_0.
 \label{eq:quadrupoleDegeneracy}
\end{equation}
Its role can be checked without relying only on the leading power at infinity. Differentiating the family with respect to $\alpha_1$ gives
\begin{equation}
 \frac{\partial X_0}{\partial\alpha_1}=-x(x-1),
 \qquad
 \frac{\partial X_2}{\partial\alpha_1}=x(x-1),
 \qquad
 \frac{\partial X_K}{\partial\alpha_1}=x^2-\frac12.
 \label{eq:quadrupoleAlphaVariation}
\end{equation}
These are precisely a homogeneous GR perturbation: $\delta H=x(x-1)=-H_2^{\rm GR}/12$ and $K_{\rm GR}[\delta H]=x^2-1/2$. Varying $q_0$ therefore changes only the amplitude of the growing tide and cannot alter the decaying response. Setting that additional growing component to zero gives
\begin{equation}
 \alpha_1=-24,
 \qquad q_0=-84.
 \label{eq:quadrupoleNoTide}
\end{equation}

The spatial decaying coefficient in \eqref{eq:quadrupoleFamily} is $-480x^{-3}$. Since
\begin{equation}
 Q_2^2(2x-1)=\frac{1}{5x^3}+\order(x^{-5}),
 \label{eq:Q2asymptotic}
\end{equation}
the spatial Regge--Wheeler branch ratio is
\begin{equation}
 \Delta(B/A)_{2}^{\rm RW,fixed}=-2400\epse.
 \label{eq:fixedQuadrupoleMetricRatio}
\end{equation}

The gauge-invariant correction obtained from \eqref{eq:RWZM}, including the dressing \eqref{eq:backgroundDressing}, is
\begin{equation}
 \delta\Psi_2^{\rm ZM}(x)
 =\frac{4\left(8x^6+6x^5-480x^3+296x^2+93x+28\right)}
 {x^4(4x+3)}.
 \label{eq:exactQuadrupoleZM}
\end{equation}
At infinity,
\begin{equation}
 \delta\Psi_2^{\rm ZM}
 =8x-\frac{480}{x^2}+\frac{656}{x^3}
 -\frac{399}{x^4}+\order(x^{-5}).
 \label{eq:quadrupoleZMInfinity}
\end{equation}
The decaying homogeneous master branch begins at $1/(5x^2)$, so the independent gauge-invariant extraction gives
\begin{equation}
 \Delta(B/A)_{2}^{\rm ZM,fixed}=-2400\epse.
 \label{eq:fixedQuadrupoleRatio}
\end{equation}
It agrees with the Regge--Wheeler spatial ratio \eqref{eq:fixedQuadrupoleMetricRatio}; the gauge-invariant statement is the value obtained from $\Psi_2^{\rm ZM}$. No logarithm appears, in agreement with \eqref{eq:quadrupoleException}.

The fixed-integer value is not the canonical electric Love number. In the canonical analytic-continuation prescription~\cite{Cano2025},
\begin{equation}
 k_2^E=28\frac{\lambda_{\rm ev}}{\Mzero^4}=448\epse.
 \label{eq:canonicalQuadrupoleLove}
\end{equation}
Both the Zerilli--Moncrief ratio \eqref{eq:fixedQuadrupoleRatio} and the canonical Love number \eqref{eq:canonicalQuadrupoleLove} are gauge invariant, but they are finite parts defined by different prescriptions. At a noninteger angular momentum, source and response powers are separated unambiguously. Taking the analytic limit to $\ell=2$ defines the canonical coefficient. If the calculation is instead specialized to the integer multipole first, local terms and homogeneous pieces can be rearranged by finite subtractions, leaving the fixed branch ratio \eqref{eq:fixedQuadrupoleRatio}. The absence of running does not remove this finite ambiguity. The value \eqref{eq:canonicalQuadrupoleLove} is quoted for the Riemann-cubic convention and mapped to the Weyl-cubic calculation through the common Ricci-flat representative in \eqref{eq:couplingMap}. Since the two Lagrangians are not literally identical off shell, this comparison assumes the accompanying field-redefinition map and the same asymptotic normalization.

This distinction is important for matching. Substituting $-2400\epse$ in a worldline Love-number relation would give the wrong sign and magnitude for the quadrupolar Wilson coefficient. The correct matching uses \eqref{eq:canonicalQuadrupoleLove}, as shown in \cref{sec:canonicalWorldline}.

\section{Global octupolar solution}
\label{sec:octupole}

The octupole is the first physical electric multipole for which the cubic interaction generates logarithmic running. It is therefore the simplest global example in which the local residue calculation, horizon regularity, asymptotic matching, and scale-dependent response can all be tested simultaneously. We construct the exact Green-function solution and use it to verify explicitly the generic result derived in \cref{sec:runningProof}.

For $\ell=3$, $L=12$ and
\begin{equation}
 H_3(x)=P_3^2(2x-1)=-60x(x-1)(2x-1).
 \label{eq:octupoleGR}
\end{equation}
The dimensionless source in \eqref{eq:scalarMasterDimensionless} becomes
\begin{equation}
 j_3(x)=\frac{120\left(456-1107x+344x^2+1104x^3-1152x^4+360x^5+9x^7-26x^8+18x^9\right)}{(x-1)^2x^7}.
 \label{eq:octupoleSource}
\end{equation}
The two homogeneous functions are $P_3^2(2x-1)$ and
\begin{multline}
 Q_3^2(2x-1)=
 -\frac{-1-10x+130x^2-240x^3+120x^4}{2x(x-1)}
 \\
 -30x(x-1)(2x-1)\log\!\left(\frac{x-1}{x}\right).
 \label{eq:Q3exact}
\end{multline}

\subsection{Green-function channels and horizon anchoring}

The $Q$-channel derivative in \eqref{eq:QchannelIntegrand} is elementary:
\begin{align}
 I_{Q,3}'={}&-4320x^5+8400x^4-5280x^3+1080x^2-86400x+319680
 \notag\\
 &-\frac{403200}{x}+\frac{49920}{x^2}+\frac{306960}{x^3}
 -\frac{242280}{x^4}+\frac{54720}{x^5}.
 \label{eq:IQ3prime}
\end{align}
A primitive is
\begin{align}
 I_{Q,3}={}&-720x^6+1680x^5-1320x^4+360x^3-43200x^2+319680x
 \notag\\
 &-403200\log x-\frac{49920}{x}-\frac{153480}{x^2}
 +\frac{80760}{x^3}-\frac{13680}{x^4}.
 \label{eq:IQ3}
\end{align}
Since $I_{Q,3}(1)=140160$, the horizon-anchored coefficient is
\begin{equation}
 C_Q(x)=I_{Q,3}(x)-140160.
 \label{eq:CQ3}
\end{equation}
It vanishes linearly at the horizon, canceling the $1/(x-1)$ singularity of $Q_3^2$.

The $P$ channel is more involved and contains a dilogarithm. Its exact horizon-normalized coefficient is given in \cref{eq:CP3global}. Near $y=x-1=0$, the separate channel coefficients behave as
\begin{align}
 C_P&=\frac{6}{y}+360y\log y+1113y+2520y^2\log y+2891y^2+\cdots,
 \\
 C_Q&=-720y+\order(y^2).
 \label{eq:octupoleChannelNearHorizon}
\end{align}
Although $P_3^2C_P$ and $Q_3^2C_Q$ separately contain logarithms in their local expansions, the full particular solution
\begin{equation}
 X_0^{\rm part}=P_3^2(2x-1)C_P(x)+Q_3^2(2x-1)C_Q(x)
 \label{eq:octupoleParticular}
\end{equation}
is analytic through the calculated orders:
\begin{equation}
 X_0^{\rm part}=-720+360y-480y^2-6840y^3+25680y^4-66480y^5+\cdots.
 \label{eq:octupoleHorizonParticular}
\end{equation}
This cancellation demonstrates explicitly that the logarithmic running at infinity is compatible with a smooth future horizon.

The complete regular solution includes one homogeneous constant,
\begin{equation}
 X_0=X_0^{\rm part}+c\,P_3^2(2x-1).
 \label{eq:octupoleRegularFamily}
\end{equation}
The reconstruction formulas are linear. Consequently, the term $cP_3^2$ in $X_0$ generates the corresponding homogeneous GR companions in $X_K$ and $X_2$ and changes only the growing tidal solution. The other metric fields follow from \eqref{eq:XKfromX0} and \eqref{eq:X2AlgebraicDimensionless}. At the horizon,
\begin{equation}
 X_0(1)=X_2(1)=-720,
 \qquad
 X_K(1)=-432+12c,
 \label{eq:octupoleHorizonData}
\end{equation}
in agreement with the generic result \eqref{eq:genericHorizonData}.

\subsection{No-tidal-renormalization condition and asymptotics}

The horizon-normalized $P$-channel coefficient approaches
\begin{equation}
 C_P(\infty)=331620-33600\pi^2.
 \label{eq:CPinfinity}
\end{equation}
Because the free constant multiplies the homogeneous $P_3^2$ solution in every reconstructed component, removing the additional tidal branch fixes
\begin{equation}
 c_{\rm no\mbox{-}tide}=33600\pi^2-331620.
 \label{eq:octupoleNoTideConstant}
\end{equation}
The resulting global metric fields have the asymptotic expansions
\begin{subequations}
\begin{align}
 X_0={}&-360x^2+240x-\frac{4320}{x^2}+\frac{18720}{x^3}
 +\frac{-28800\log x+186240/7}{x^4}
 +\order\!\left(\frac{\log x}{x^5}\right),
 \\
 X_K={}&240x^2-72-\frac{8640}{x^2}-\frac{7200}{x^3}
 +\frac{28800\log x-155160/7}{x^4}
 +\order\!\left(\frac{\log x}{x^5}\right),
 \\
 X_2={}&360x^2-240x+\frac{12960}{x^2}-\frac{11520}{x^3}
 +\frac{28800\log x-428160/7}{x^4}
 +\order\!\left(\frac{\log x}{x^5}\right).
 \label{eq:octupoleMetricAsymptotics}
\end{align}
\end{subequations}
The metric logarithmic coefficients agree with the generic formula: $\betahat_{X_0}^{(3)}=-403200$ and $\betahat_{X_K}^{(3)}=\betahat_{X_2}^{(3)}=403200$.

The decaying master branch satisfies
\begin{equation}
 \Psi_{Q,3}^{\rm ZM}=\frac{1}{42x^3}+\order(x^{-4}),
 \label{eq:Q3MasterAsymptotic}
\end{equation}
and the complete correction contains
\begin{equation}
 \delta\Psi_3^{\rm ZM}\supset\frac{9600\epse\log x}{x^3}.
 \label{eq:octupoleRawMasterLog}
\end{equation}
The ratio $9600/(1/42)$ reproduces $\beta_3^{\rm ZM}=403200\epse$. The finite coefficient depends on the subtraction length and is recorded in \cref{app:octupoleGlobal}; its evolution is governed by \eqref{eq:ZMRGlaw}. The $\pi^2$ in the no-tide constant \eqref{eq:octupoleNoTideConstant} cancels against terms from the global Green-function channels in several asymptotic metric coefficients, while the logarithmic coefficient remains fixed by the local residue.

\section{Canonical normalization and finite-size worldline running}
\label{sec:canonicalWorldline}

The coefficient in \eqref{eq:betaZM} refers to the decaying Zerilli--Moncrief branch. Canonical Love numbers are instead normalized through the asymptotic Teukolsky response. We now derive the conversion rather than fixing it from a low-multipole comparison, and then use the same convention to define the finite-size worldline coupling.

\subsection{From Zerilli--Moncrief to the canonical electric response}

Direct substitution of the growing GR solution in \eqref{eq:RWZM} gives
\begin{equation}
 \Psi_{P,\ell}^{\rm ZM}
 \sim \mathcal A_\ell x^{\ell+1},
 \qquad
 \mathcal A_\ell
 =-\frac{2(2\ell)!\,\ell(\ell-1)}
 {(\ell!)^2(\ell+1)(\ell+2)}.
 \label{eq:ZMgrowingNormalization}
\end{equation}
Together with \eqref{eq:ZMdecayingNormalization}, the logarithmic asymptotic solution can therefore be written as
\begin{equation}
 \delta\Psi_\ell^{\rm ZM}
 \sim \mathcal A_\ell x^{\ell+1}
 +a_\ell^{\log}x^{-\ell}\log x+\cdots,
 \qquad
 a_\ell^{\log}=\mathcal N_\ell\beta_\ell^{\rm ZM}.
 \label{eq:ZMrawAsymptotic}
\end{equation}
The overall source amplitude in \eqref{eq:ZMrawAsymptotic} is displayed only to track normalization; the physical perturbation remains linear in an arbitrary tidal amplitude.

Let $\nu=(\ell-1)(\ell+2)/2$. At zero frequency, the Chandrasekhar transformation can be written as~\cite{Chandrasekhar1975}
\begin{equation}
 \nu(\nu+1)\Psi_\ell^{\rm ZM}
 =\left[\nu(\nu+1)
 +\frac{9\Mzero^2(r-2\Mzero)}{r^2(\nu r+3\Mzero)}\right]\psi_{\rm RW}
 +3\Mzero f\frac{\dd\psi_{\rm RW}}{\dd r}.
 \label{eq:ZerilliToRWStatic}
\end{equation}
For either power-law branch, the correction terms on the right are one radial power below the leading algebraic term. Thus $\Psi_\ell^{\rm ZM}$ and $\psi_{\rm RW}$ have the same leading source--response ratio; any common overall normalization cancels. The Regge--Wheeler function is related to the static spin-$-2$ Teukolsky variable by~\cite{Cano2025}
\begin{equation}
 \psi_{-2}\propto\frac{f}{8}
 \left[(6\Mzero-Lr)\psi_{\rm RW}
 +2r(3\Mzero-r)\frac{\dd\psi_{\rm RW}}{\dd r}\right].
 \label{eq:RWtoTeukolskyStatic}
\end{equation}
Only the relative normalization of the two asymptotic branches is needed. Acting with \eqref{eq:RWtoTeukolskyStatic} gives
\begin{align}
 r^{\ell+1}&\longmapsto
 -(\ell+1)(\ell+2)r^{\ell+2}+\text{subleading powers},
 \\
 r^{-\ell}\log r&\longmapsto
 -\ell(\ell-1)r^{1-\ell}\log r
 +\text{nonlogarithmic terms}.
 \label{eq:TeukolskyBranchMap}
\end{align}

In the canonical convention of Ref.~\cite{Cano2025}, the logarithmic response in the Teukolsky variable is written as
\begin{equation}
 \psi_{-2}\sim A_\ell^{\rm T}r^{\ell+2}
 +2A_\ell^{\rm T}\Mzero^{2\ell+1}
 \beta_\ell^+\,r^{1-\ell}\log\!\left(\frac{r}{\Mzero}\right)+\cdots.
 \label{eq:canonicalTeukolskyConvention}
\end{equation}
Since $x=r/(2\Mzero)$, the ratio of the logarithmic and growing coefficients obtained from \eqref{eq:ZMrawAsymptotic} and \eqref{eq:TeukolskyBranchMap} obeys
\begin{equation}
 2\beta_\ell^+
 =2^{2\ell+1}
 \frac{\ell(\ell-1)}{(\ell+1)(\ell+2)}
 \frac{a_\ell^{\log}}{\mathcal A_\ell}.
 \label{eq:bridgeIntermediate}
\end{equation}
Substitution of \eqref{eq:ZMgrowingNormalization} yields
\begin{equation}
 \boxed{
 \beta_\ell^+
 =-\frac{2^{2\ell-1}(\ell!)^2}{(2\ell)!}\,a_\ell^{\log}}
 \label{eq:rawLogConversion}
\end{equation}
and, using $a_\ell^{\log}=\mathcal N_\ell\beta_\ell^{\rm ZM}$,
\begin{equation}
 \beta_\ell^+=\mathcal C_\ell^{+\leftarrow{\rm ZM}}\beta_\ell^{\rm ZM},
 \qquad
 \mathcal C_\ell^{+\leftarrow{\rm ZM}}
 =-\frac{2^{2\ell-1}(\ell!)^4(\ell+1)(\ell+2)}
 {(\ell-1)\ell(2\ell+1)[(2\ell)!]^2}.
 \label{eq:canonicalBridge}
\end{equation}
Thus the factorial conversion is fixed by the asymptotic transformations of the source and response branches, not by matching one numerical multipole.

For the octupole, $a_3^{\log}=9600\epse$, and \eqref{eq:rawLogConversion} gives
\begin{equation}
 \beta_3^+=-15360\epse
 =-960\frac{\lambda_{\rm ev}}{\Mzero^4}.
 \label{eq:canonicalOctupoleBeta}
\end{equation}
The general result of Ref.~\cite{Cano2025} is
\begin{equation}
 \beta_\ell^+
 =-\frac{7}{6}\frac{\lambda_{\rm ev}}{\Mzero^4}F_\ell(L-4),
 \qquad
 F_\ell=4^{\ell-3}
 \frac{(\ell!)^4(\ell-2)(\ell-1)\ell(\ell+1)^3(\ell+2)^3(\ell+3)}
 {(2\ell+1)[(2\ell)!]^2}.
 \label{eq:CanoBeta}
\end{equation}
Equations~\eqref{eq:betaZM}, \eqref{eq:canonicalBridge}, and \eqref{eq:couplingMap} reduce algebraically to \eqref{eq:CanoBeta}. The bridge concerns the logarithmic coefficient. At $\ell=2$ both beta functions vanish, so it cannot convert the finite fixed-integer ratio into the analytically continued Love number; that finite distinction was discussed in \cref{sec:quadrupole}.

\subsection{Finite-size worldline operators}

Let $u^\mu$ be the worldline four-velocity and
\begin{equation}
 h_{\mu\nu}=g_{\mu\nu}+u_\mu u_\nu,
 \qquad
 E_{ab}=C_{a\mu b\nu}u^\mu u^\nu,
 \qquad
 D_a=h_a{}^\mu\nabla_\mu.
 \label{eq:worldlineGeometricDefinitions}
\end{equation}
Angle brackets on spatial indices denote the symmetric trace-free projection with $h_{ab}$. Define the electric STF multipole
\begin{equation}
 \mathcal E_{a_1\cdots a_\ell}
 \equiv D_{\langle a_1}\cdots D_{a_{\ell-2}}
 E_{a_{\ell-1}a_\ell\rangle}.
 \label{eq:electricSTFMultipole}
\end{equation}
The finite-size action convention used here is
\begin{equation}
 S_{\rm fs}^{(\ell)}
 =\frac{1}{4\pi G}\frac{\lambda_\ell^E}{2\ell!}
 \int\dd\tau\,
 \mathcal E_{a_1\cdots a_\ell}
 \mathcal E^{a_1\cdots a_\ell}
 \equiv c_{E,\ell}^{\rm fs}
 \int\dd\tau\,
 \mathcal E_{a_1\cdots a_\ell}
 \mathcal E^{a_1\cdots a_\ell}.
 \label{eq:worldlineOperator}
\end{equation}
This definition avoids using $L$ as a multi-index, since $L$ already denotes $\ell(\ell+1)$ in the metric calculation.

The canonical dimensionless Love number is related to the coefficient by
\begin{equation}
 c_{E,\ell}^{\rm fs}
 =\frac{\lambda_\ell^E}{8\pi G\,\ell!}
 =\frac{2^\ell k_\ell^E R_{\rm ref}^{2\ell+1}}
 {G(2\ell)!}.
 \label{eq:LoveToWilson}
\end{equation}
Here $R_{\rm ref}$ is the reference length used to make $k_\ell^E$ dimensionless; it is not a geometric surface radius. For the black-hole convention compared with Ref.~\cite{Cano2025}, we choose
\begin{equation}
 R_{\rm ref}=\Mzero=\frac{\rs}{2}.
 \label{eq:referenceLengthChoice}
\end{equation}

Using the same subtraction length $r_0$ as in \eqref{eq:ZMresponseParametrization}, the finite-size coefficient obeys
\begin{equation}
 r_0\frac{\dd c_{E,\ell}^{\rm fs}}{\dd r_0}
 =\frac{2^\ell\Mzero^{2\ell+1}}{G(2\ell)!}\,\beta_\ell^+.
 \label{eq:WilsonBetaCompact}
\end{equation}
Equivalently, with $\mu_{\rm R}=r_0^{-1}$, the right-hand side changes sign. In terms of $\epse$ and $\rs$, \eqref{eq:WilsonBetaCompact} becomes
\begin{align}
 r_0\frac{\dd c_{E,\ell}^{\rm fs}}{\dd r_0}
 =-\frac{7\,2^{\ell-4}}{3G}
 \frac{(\ell-2)(\ell-1)\ell(\ell+1)^3(\ell+2)^3(\ell+3)}{2\ell+1}
  \frac{(\ell^2+\ell-4)(\ell!)^4}{[(2\ell)!]^3}
 \epse\rs^{2\ell+1}.
 \label{eq:WilsonBetaExplicit}
\end{align}
The dimension is $[c_{E,\ell}^{\rm fs}]=({\rm length})^{2\ell-1}$.

Define the length-scale coefficient $\gamma_\ell^{(r)}$ by
\begin{equation}
 r_0\frac{\dd c_{E,\ell}^{\rm fs}}{\dd r_0}
 =\gamma_\ell^{(r)}\frac{\epse\rs^{2\ell+1}}{G}.
 \label{eq:gammaDefinition}
\end{equation}
\begin{table}[t]
\centering
\caption{Canonical electric beta functions and length-scale running of the finite-size worldline coefficients.}
\label{tab:canonicalAndWilson}
\small
\begin{tabular}{@{}cccc@{}}
\toprule
$\ell$ & $\betahat_\ell$ & $\beta_\ell^+/\epse$ & $\gamma_\ell^{(r)}$ \\
\midrule
2 & $0$ & $0$ & $0$ \\
3 & $403{,}200$ & $-15{,}360$ & $-4/3$ \\
4 & $16{,}934{,}400$ & $-122{,}880$ & $-2/21$ \\
5 & $256{,}838{,}400$ & $-13{,}045{,}760/33$ & $-91/53{,}460$ \\
6 & $2{,}252{,}275{,}200$ & $-1{,}220{,}280{,}320/1573$ & $-266/21{,}023{,}145$ \\
\bottomrule
\end{tabular}
\end{table}

At $\ell=2$, \eqref{eq:LoveToWilson} and \eqref{eq:canonicalQuadrupoleLove} give
\begin{equation}
 c_{E,2}^{\rm fs}
 =\frac{7}{3G}\epse\rs^5
 =\frac{14}{3}\lambda_{\rm ev}m_0,
 \qquad m_0\equiv\frac{\Mzero}{G}.
 \label{eq:quadrupoleWilson}
\end{equation}
For the quadrupolar Riemann-cubic representative, the explicit point-source matching of Ref.~\cite{WangLehnerMicolSturani2026} finds $c_{E,2}^{\rm pp}=0$, so $c_{E,2}^{\rm full}=c_{E,2}^{\rm fs}$ in that convention. At $\ell=3$,
\begin{equation}
 r_0\frac{\dd c_{E,3}^{\rm fs}}{\dd r_0}
 =-\frac{4}{3G}\epse\rs^7
 =-\frac{32}{3G}\lambda_{\rm ev}\Mzero^3.
 \label{eq:octupoleWilsonBeta}
\end{equation}

For higher multipoles the exterior response fixes the finite-size term, but a complete distributional matching may also contain
\begin{equation}
 c_{E,\ell}^{\rm full}=c_{E,\ell}^{\rm fs}+c_{E,\ell}^{\rm pp}.
 \label{eq:WilsonDecomposition}
\end{equation}
No generic $\ell\geq3$ point-source calculation is performed here. Equation~\eqref{eq:WilsonBetaExplicit} is therefore the running of the finite-size coefficient, not a claim about the complete distributional Wilson coefficient.

\section{Discussion and conclusions}
\label{sec:discussion}

The calculation gives a metric description of the static electric response for the full Schwarzschild multipole tower in parity-even cubic gravity. The low-degree dependence on $L=\ell(\ell+1)$ is already visible at the action level: integration by parts on the sphere leaves at most one power of $L$ in the Einstein terms and two in the cubic term. The projected multipoles determine those polynomials, and the remaining integer multipoles test them independently.

The reduced equations preserve the constraint structure of static even-parity perturbations. Two first-order equations and one algebraic relation reconstruct the three metric functions. Eliminating $X_K$ then produces the GR tidal operator with a new source. This separation is useful physically: the higher-curvature interaction does not introduce an additional homogeneous mode at the order considered; it drives the familiar growing and decaying GR branches.

The logarithm is controlled by a single Laurent coefficient in the decaying Green-function channel. After the GR recurrence is imposed, that coefficient factorizes as $(L-6)(L-4)L^2(L-2)^2$. The horizon residue vanishes independently. The absence of running at $\ell=2$ is therefore traced to the source that excites the decaying branch, rather than to a cancellation imposed later by a gauge choice or by the canonical normalization. Every electric multipole with $\ell\geq3$ is resonant.

Passing to the Zerilli--Moncrief variable confirms that the logarithm belongs to the decaying, gauge-invariant branch. The derivative in the master-variable map adds a finite local term but leaves its logarithmic coefficient unchanged. The subsequent conversion to the canonical Teukolsky response follows from the zero-frequency Chandrasekhar map and from the different asymptotic weights of the two Teukolsky branches. This derivation reproduces the canonical beta function of Ref.~\cite{Cano2025} without using a fixed multipole to calibrate the normalization.

The quadrupole and octupole illuminate different parts of the result. At $\ell=2$, the free regular horizon datum is shown explicitly to be a homogeneous growing tide. Removing it leaves the fixed-integer Zerilli--Moncrief ratio $-2400\epse$. The canonical value $k_2^E=448\epse$ is different because it is defined by analytic continuation before the integer-multipole degeneracy is taken. At $\ell=3$, the two Green-function channels contain separate horizon logarithms, but these cancel in the full solution; the surviving logarithm at infinity has precisely the coefficient predicted by the generic residue.

The worldline translation has been stated with two qualifications. First, the scale $r_0$ in the exterior logarithm is a subtraction length; the anomalous dimension written with an energy scale $\mu_{\rm R}=r_0^{-1}$ has the opposite sign. Second, the exterior solution determines the finite-size coefficient. A generic point-particle counterterm for $\ell\geq3$ would require a separate distributional matching and is not inferred here.

The canonical beta functions themselves were previously obtained through the modified Teukolsky equation. The additional content of the present approach is the radial action, the constrained metric system, the full metric reconstruction, the source-level origin of the factor $L-6$, and the explicit relation between the fixed-integer and canonical quadrupolar prescriptions. These ingredients are useful when an observable requires metric components rather than only a curvature master variable.

The analysis is restricted to static electric perturbations of the parity-even cubic interaction. Magnetic and parity-mixing sectors, rotation, finite frequency, and the generic point-source matching remain outside its scope. Within the stated approximation, however, the action, metric equations, Green-function residue, gauge-invariant running, low-multipole global solutions, canonical conversion, and finite-size worldline running form a closed and mutually consistent calculation.

\section*{Data and code availability}
Ancillary files included with this submission contain the closed radial actions, the first-order matrix and sources, the scalar source, the global octupolar solution, and the response formulas listed in \cref{app:validation}. The driver \texttt{VerifyMainResults.m} loads these machine-readable expressions and checks the stored matrix and reconstruction formula, the low-multipole response anchors, and the generic normalization identities. All symbolic inputs are exact; no floating-point data are required for these checks.

\section*{Acknowledgments}
This work was partially supported by the Brazilian agencies CAPES, CNPq, and FAPEMA. E.O.S. acknowledges support from grants CNPq/306308/2022-3, FAPEMA/UNIVERSAL-06395/22, and CAPES/Finance Code 001.

\appendix

\section{Complete arbitrary-\texorpdfstring{$L$}{L} radial-action coefficients}
\label{app:actions}

This appendix gives the complete coefficient representation \eqref{eq:actionCoefficientRepresentation}. Primes in \cref{tab:action_EH0_coefficients,tab:action_EH1_coefficients,tab:action_C3_coefficients} denote derivatives with respect to $r$, as in the original radial action. The three tables use the same two-column convention: the monomial is listed in the first column and its exact coefficient in the second, with both entries centered horizontally and vertically. Terms related by integrations by parts have not been removed; the tables reproduce the exact representative used to derive the Euler--Lagrange equations. Adding a total radial derivative changes individual coefficients but leaves all equations in the main text invariant.

\subsection{Einstein--Hilbert action on Schwarzschild}

\begingroup
\small
\renewcommand{\arraystretch}{1.18}
\setlength{\tabcolsep}{8pt}
\setlength{\LTleft}{\fill}
\setlength{\LTright}{\fill}
\begin{longtable}{@{}M{0.18\textwidth}M{0.76\textwidth}@{}}
\caption{Coefficients of the Einstein--Hilbert radial action on the Schwarzschild background, $\mathcal L_{\rm EH}^{(0)}$, in the representation \eqref{eq:actionCoefficientRepresentation}.}\label{tab:action_EH0_coefficients}\\
\toprule
\textbf{Monomial $\mathcal M$} & \textbf{Coefficient $c_{\mathcal M}(r,r_s;L)$} \\
\midrule
\endfirsthead
\multicolumn{2}{c}{\itshape Table \thetable\ continued}\\[-0.15em]
\toprule
\textbf{Monomial $\mathcal M$} & \textbf{Coefficient $c_{\mathcal M}(r,r_s;L)$} \\
\midrule
\endhead
\midrule
\multicolumn{2}{r@{}}{\itshape continued on next page}\\
\endfoot
\bottomrule
\endlastfoot
\actiontablesection{Algebraic terms}
$\displaystyle H_2\,K$ & $\displaystyle \frac{L + 2}{2}$ \\[1.5mm]
$\displaystyle (H_2)^{2}$ & $\displaystyle -1$ \\[1.5mm]
$\displaystyle H_0\,K$ & $\displaystyle \frac{L - 2}{2}$ \\[1.5mm]
$\displaystyle H_0\,H_2$ & $\displaystyle \frac{L + 2}{2}$ \\[1.5mm]
\actiontablesection{Terms with first derivatives}
$\displaystyle (K')^{2}$ & $\displaystyle \frac{r \left(r - r_s\right)}{2}$ \\[1.5mm]
$\displaystyle H_2'\,K'$ & $\displaystyle r \left(r - r_s\right)$ \\[1.5mm]
$\displaystyle H_0'\,K'$ & $\displaystyle - r \left(r - r_s\right)$ \\[1.5mm]
$\displaystyle H_0'\,H_2'$ & $\displaystyle \frac{r \left(r - r_s\right)}{2}$ \\[1.5mm]
$\displaystyle (H_0')^{2}$ & $\displaystyle \frac{r \left(r - r_s\right)}{2}$ \\[1.5mm]
$\displaystyle K\,H_2'$ & $\displaystyle \frac{4 r - 3 r_s}{2}$ \\[1.5mm]
$\displaystyle K\,H_0'$ & $\displaystyle - \frac{4 r - r_s}{2}$ \\[1.5mm]
$\displaystyle H_2\,K'$ & $\displaystyle 3 r - 2 r_s$ \\[1.5mm]
$\displaystyle H_2\,H_2'$ & $\displaystyle - \frac{3 \left(4 r - 3 r_s\right)}{4}$ \\[1.5mm]
$\displaystyle H_2\,H_0'$ & $\displaystyle \frac{4 r - r_s}{4}$ \\[1.5mm]
$\displaystyle H_0\,K'$ & $\displaystyle - 3 r + 2 r_s$ \\[1.5mm]
$\displaystyle H_0\,H_2'$ & $\displaystyle \frac{4 r - 3 r_s}{4}$ \\[1.5mm]
$\displaystyle H_0\,H_0'$ & $\displaystyle \frac{4 r - r_s}{4}$ \\[1.5mm]
\actiontablesection{Terms with second derivatives}
$\displaystyle K\,H_0''$ & $\displaystyle - r \left(r - r_s\right)$ \\[1.5mm]
$\displaystyle H_2\,K''$ & $\displaystyle r \left(r - r_s\right)$ \\[1.5mm]
$\displaystyle H_2\,H_0''$ & $\displaystyle \frac{r \left(r - r_s\right)}{2}$ \\[1.5mm]
$\displaystyle H_0\,K''$ & $\displaystyle - r \left(r - r_s\right)$ \\[1.5mm]
$\displaystyle H_0\,H_0''$ & $\displaystyle \frac{r \left(r - r_s\right)}{2}$ \\[1.5mm]
\end{longtable}
\endgroup

\subsection{Einstein--Hilbert action on the corrected background}

\begingroup
\small
\renewcommand{\arraystretch}{1.18}
\setlength{\tabcolsep}{8pt}
\setlength{\LTleft}{\fill}
\setlength{\LTright}{\fill}
\begin{longtable}{@{}M{0.18\textwidth}M{0.76\textwidth}@{}}
\caption{Coefficients of the Einstein--Hilbert radial action induced by the $\order(\epse)$ background correction, $\mathcal L_{\rm EH}^{(1)}$, in the representation \eqref{eq:actionCoefficientRepresentation}.}\label{tab:action_EH1_coefficients}\\
\toprule
\textbf{Monomial $\mathcal M$} & \textbf{Coefficient $c_{\mathcal M}(r,r_s;L)$} \\
\midrule
\endfirsthead
\multicolumn{2}{c}{\itshape Table \thetable\ continued}\\[-0.15em]
\toprule
\textbf{Monomial $\mathcal M$} & \textbf{Coefficient $c_{\mathcal M}(r,r_s;L)$} \\
\midrule
\endhead
\midrule
\multicolumn{2}{r@{}}{\itshape continued on next page}\\
\endfoot
\bottomrule
\endlastfoot
\actiontablesection{Algebraic terms}
$\displaystyle H_2\,K$ & $\displaystyle - \frac{3 r_s^{6} \left(L r + 2 r + 2 r_s\right)}{r^{7}}$ \\[1.5mm]
$\displaystyle (H_2)^{2}$ & $\displaystyle \frac{3 r_s^{6} \left(4 r + 3 r_s\right)}{2 r^{7}}$ \\[1.5mm]
$\displaystyle H_0\,K$ & $\displaystyle - \frac{3 r_s^{6} \left(L r - 2 r - 2 r_s\right)}{r^{7}}$ \\[1.5mm]
$\displaystyle H_0\,H_2$ & $\displaystyle - \frac{3 r_s^{6} \left(L r + 2 r + r_s\right)}{r^{7}}$ \\[1.5mm]
$\displaystyle (H_0)^{2}$ & $\displaystyle - \frac{3 r_s^{7}}{2 r^{7}}$ \\[1.5mm]
\actiontablesection{Terms with first derivatives}
$\displaystyle (K')^{2}$ & $\displaystyle - \frac{r_s \left(r - r_s\right)^{2} \left(r^{4} + 2 r^{3} r_s + 3 r^{2} r_s^{2} + 4 r r_s^{3} + 5 r_s^{4}\right)}{r^{5}}$ \\[1.5mm]
$\displaystyle H_2'\,K'$ & $\displaystyle - \frac{2 r_s \left(r - r_s\right)^{2} \left(r^{4} + 2 r^{3} r_s + 3 r^{2} r_s^{2} + 4 r r_s^{3} + 5 r_s^{4}\right)}{r^{5}}$ \\[1.5mm]
$\displaystyle H_0'\,K'$ & $\displaystyle \frac{2 r_s \left(r - r_s\right)^{2} \left(r^{4} + 2 r^{3} r_s + 3 r^{2} r_s^{2} + 4 r r_s^{3} + 5 r_s^{4}\right)}{r^{5}}$ \\[1.5mm]
$\displaystyle H_0'\,H_2'$ & $\displaystyle - \frac{r_s \left(r - r_s\right)^{2} \left(r^{4} + 2 r^{3} r_s + 3 r^{2} r_s^{2} + 4 r r_s^{3} + 5 r_s^{4}\right)}{r^{5}}$ \\[1.5mm]
$\displaystyle (H_0')^{2}$ & $\displaystyle - \frac{r_s \left(r - r_s\right)^{2} \left(r^{4} + 2 r^{3} r_s + 3 r^{2} r_s^{2} + 4 r r_s^{3} + 5 r_s^{4}\right)}{r^{5}}$ \\[1.5mm]
$\displaystyle K\,H_2'$ & $\displaystyle - \frac{3 r_s \left(r - r_s\right) \left(r^{5} + r^{4} r_s + r^{3} r_s^{2} + r^{2} r_s^{3} + r r_s^{4} - r_s^{5}\right)}{r^{6}}$ \\[1.5mm]
$\displaystyle K\,H_0'$ & $\displaystyle \frac{r_s \left(r - r_s\right) \left(r^{5} + r^{4} r_s + r^{3} r_s^{2} + r^{2} r_s^{3} + r r_s^{4} + 67 r_s^{5}\right)}{r^{6}}$ \\[1.5mm]
$\displaystyle H_2\,K'$ & $\displaystyle - \frac{4 r_s \left(r - r_s\right) \left(r^{5} + r^{4} r_s + r^{3} r_s^{2} + r^{2} r_s^{3} + r r_s^{4} + 10 r_s^{5}\right)}{r^{6}}$ \\[1.5mm]
$\displaystyle H_2\,H_2'$ & $\displaystyle \frac{9 r_s \left(r - r_s\right) \left(r^{5} + r^{4} r_s + r^{3} r_s^{2} + r^{2} r_s^{3} + r r_s^{4} - r_s^{5}\right)}{2 r^{6}}$ \\[1.5mm]
$\displaystyle H_2\,H_0'$ & $\displaystyle - \frac{r_s \left(r - r_s\right) \left(r^{5} + r^{4} r_s + r^{3} r_s^{2} + r^{2} r_s^{3} + r r_s^{4} + 67 r_s^{5}\right)}{2 r^{6}}$ \\[1.5mm]
$\displaystyle H_0\,K'$ & $\displaystyle \frac{4 r_s \left(r - r_s\right) \left(r^{5} + r^{4} r_s + r^{3} r_s^{2} + r^{2} r_s^{3} + r r_s^{4} + 10 r_s^{5}\right)}{r^{6}}$ \\[1.5mm]
$\displaystyle H_0\,H_2'$ & $\displaystyle - \frac{3 r_s \left(r - r_s\right) \left(r^{5} + r^{4} r_s + r^{3} r_s^{2} + r^{2} r_s^{3} + r r_s^{4} - r_s^{5}\right)}{2 r^{6}}$ \\[1.5mm]
$\displaystyle H_0\,H_0'$ & $\displaystyle - \frac{r_s \left(r - r_s\right) \left(r^{5} + r^{4} r_s + r^{3} r_s^{2} + r^{2} r_s^{3} + r r_s^{4} + 67 r_s^{5}\right)}{2 r^{6}}$ \\[1.5mm]
\actiontablesection{Terms with second derivatives}
$\displaystyle K\,H_0''$ & $\displaystyle \frac{2 r_s \left(r - r_s\right)^{2} \left(r^{4} + 2 r^{3} r_s + 3 r^{2} r_s^{2} + 4 r r_s^{3} + 5 r_s^{4}\right)}{r^{5}}$ \\[1.5mm]
$\displaystyle H_2\,K''$ & $\displaystyle - \frac{2 r_s \left(r - r_s\right)^{2} \left(r^{4} + 2 r^{3} r_s + 3 r^{2} r_s^{2} + 4 r r_s^{3} + 5 r_s^{4}\right)}{r^{5}}$ \\[1.5mm]
$\displaystyle H_2\,H_0''$ & $\displaystyle - \frac{r_s \left(r - r_s\right)^{2} \left(r^{4} + 2 r^{3} r_s + 3 r^{2} r_s^{2} + 4 r r_s^{3} + 5 r_s^{4}\right)}{r^{5}}$ \\[1.5mm]
$\displaystyle H_0\,K''$ & $\displaystyle \frac{2 r_s \left(r - r_s\right)^{2} \left(r^{4} + 2 r^{3} r_s + 3 r^{2} r_s^{2} + 4 r r_s^{3} + 5 r_s^{4}\right)}{r^{5}}$ \\[1.5mm]
$\displaystyle H_0\,H_0''$ & $\displaystyle - \frac{r_s \left(r - r_s\right)^{2} \left(r^{4} + 2 r^{3} r_s + 3 r^{2} r_s^{2} + 4 r r_s^{3} + 5 r_s^{4}\right)}{r^{5}}$ \\[1.5mm]
\end{longtable}
\endgroup

\subsection{Cubic-curvature contribution}
\begingroup
\small
\renewcommand{\arraystretch}{1.18}
\setlength{\tabcolsep}{8pt}
\setlength{\LTleft}{\fill}
\setlength{\LTright}{\fill}
\setlength{\LTcapwidth}{\textwidth}

\begin{longtable}{@{}M{0.18\textwidth}M{0.76\textwidth}@{}}
\caption{Coefficients of the parity-even cubic-curvature radial action, $\mathcal L_{C^3}^{(0)}$, in the representation \eqref{eq:actionCoefficientRepresentation}.}\label{tab:action_C3_coefficients}\\
\toprule
\textbf{Monomial $\mathcal M$} & \textbf{Coefficient $c_{\mathcal M}(r,r_s;L)$} \\
\midrule
\endfirsthead
\multicolumn{2}{c}{\itshape Table \thetable\ continued}\\[-0.15em]
\toprule
\textbf{Monomial $\mathcal M$} & \textbf{Coefficient $c_{\mathcal M}(r,r_s;L)$} \\
\midrule
\endhead
\midrule
\multicolumn{2}{r@{}}{\itshape continued on next page}\\
\endfoot
\bottomrule
\endlastfoot
\actiontablesection{Algebraic terms}
$\displaystyle (K)^{2}$ & $\displaystyle \frac{r_s \left(L - 2\right)^{2}}{r^{5}}$ \\[1.5mm]
$\displaystyle H_2\,K$ & $\displaystyle - \frac{r_s \left(L^{2} r^{2} - 6 L r^{2} + 9 L r r_s + 8 r^{2} - 30 r r_s + 30 r_s^{2}\right)}{r^{7}}$ \\[1.5mm]
$\displaystyle (H_2)^{2}$ & $\displaystyle - \frac{r_s \left(\splitfrac{4 L^{2} r^{3} - 4 L^{2} r^{2} r_s - 8 L r^{3} - 16 L r^{2} r_s + 21 L r r_s^{2}}{- 32 r^{3} + 272 r^{2} r_s - 660 r r_s^{2} + 420 r_s^{3}}\right)}{8 r^{7} \left(r - r_s\right)}$ \\[1.5mm]
$\displaystyle H_0\,K$ & $\displaystyle - \frac{r_s \left(L^{2} r^{2} - 2 L r^{2} - 3 L r r_s + 6 r r_s - 6 r_s^{2}\right)}{r^{7}}$ \\[1.5mm]
$\displaystyle H_0\,H_2$ & $\displaystyle \frac{r_s \left(8 L^{2} r^{3} - 8 L^{2} r^{2} r_s - 32 L r^{3} + 56 L r^{2} r_s - 27 L r r_s^{2} + 24 r^{2} r_s - 84 r r_s^{2} + 60 r_s^{3}\right)}{4 r^{7} \left(r - r_s\right)}$ \\[1.5mm]
$\displaystyle (H_0)^{2}$ & $\displaystyle - \frac{r_s \left(4 L^{2} r^{3} - 4 L^{2} r^{2} r_s - 24 L r^{3} + 48 L r^{2} r_s - 27 L r r_s^{2} + 12 r r_s^{2} - 12 r_s^{3}\right)}{8 r^{7} \left(r - r_s\right)}$ \\[1.5mm]
\actiontablesection{Terms with first derivatives}
$\displaystyle (K')^{2}$ & $\displaystyle \frac{r_s \left(3 L r^{2} - 3 L r r_s - 12 r r_s + 14 r_s^{2}\right)}{2 r^{5}}$ \\[1.5mm]
$\displaystyle H_2'\,K'$ & $\displaystyle - \frac{r_s^{2} \left(5 r - 6 r_s\right)}{r^{5}}$ \\[1.5mm]
$\displaystyle (H_2')^{2}$ & $\displaystyle \frac{r_s \left(2 r - 3 r_s\right)^{2}}{4 r^{5}}$ \\[1.5mm]
$\displaystyle H_0'\,K'$ & $\displaystyle - \frac{r_s \left(3 L r^{2} - 3 L r r_s - 5 r r_s + 8 r_s^{2}\right)}{r^{5}}$ \\[1.5mm]
$\displaystyle H_0'\,H_2'$ & $\displaystyle - \frac{r_s \left(4 r^{2} - 22 r r_s + 21 r_s^{2}\right)}{2 r^{5}}$ \\[1.5mm]
$\displaystyle (H_0')^{2}$ & $\displaystyle \frac{r_s \left(6 L r^{2} - 6 L r r_s + 4 r^{2} - 8 r r_s + 13 r_s^{2}\right)}{4 r^{5}}$ \\[1.5mm]
$\displaystyle K\,K'$ & $\displaystyle \frac{2 r_s^{2} \left(L - 2\right)}{r^{5}}$ \\[1.5mm]
$\displaystyle K\,H_2'$ & $\displaystyle - \frac{r_s \left(2 r - 3 r_s\right) \left(L r - 2 r + 3 r_s\right)}{r^{6}}$ \\[1.5mm]
$\displaystyle K\,H_0'$ & $\displaystyle \frac{r_s \left(2 r - 5 r_s\right) \left(L r - 2 r + 3 r_s\right)}{r^{6}}$ \\[1.5mm]
$\displaystyle H_2\,K'$ & $\displaystyle - \frac{r_s \left(6 L r^{2} - 7 L r r_s - 8 r r_s + 30 r_s^{2}\right)}{2 r^{6}}$ \\[1.5mm]
$\displaystyle H_2\,H_2'$ & $\displaystyle \frac{r_s \left(2 r - 3 r_s\right) \left(L r - 4 r + 21 r_s\right)}{2 r^{6}}$ \\[1.5mm]
$\displaystyle H_2\,H_0'$ & $\displaystyle \frac{r_s \left(4 L r^{2} - 4 L r r_s + 8 r^{2} - 50 r r_s + 75 r_s^{2}\right)}{2 r^{6}}$ \\[1.5mm]
$\displaystyle H_0\,K'$ & $\displaystyle \frac{r_s \left(6 L r^{2} - 11 L r r_s + 6 r_s^{2}\right)}{2 r^{6}}$ \\[1.5mm]
$\displaystyle H_0\,H_2'$ & $\displaystyle \frac{r_s \left(2 r - 3 r_s\right) \left(L r - 3 r_s\right)}{2 r^{6}}$ \\[1.5mm]
$\displaystyle H_0\,H_0'$ & $\displaystyle - \frac{r_s \left(8 L r^{2} - 14 L r r_s + 6 r r_s - 15 r_s^{2}\right)}{2 r^{6}}$ \\[1.5mm]
\actiontablesection{Terms with second derivatives}
$\displaystyle (K'')^{2}$ & $\displaystyle \frac{r_s \left(r - r_s\right)^{2}}{r^{3}}$ \\[1.5mm]
$\displaystyle H_0''\,K''$ & $\displaystyle - \frac{2 r_s \left(r - r_s\right)^{2}}{r^{3}}$ \\[1.5mm]
$\displaystyle (H_0'')^{2}$ & $\displaystyle \frac{r_s \left(r - r_s\right)^{2}}{r^{3}}$ \\[1.5mm]
$\displaystyle K'\,K''$ & $\displaystyle \frac{2 r_s^{2} \left(r - r_s\right)}{r^{4}}$ \\[1.5mm]
$\displaystyle K'\,H_0''$ & $\displaystyle - \frac{2 r_s^{2} \left(r - r_s\right)}{r^{4}}$ \\[1.5mm]
$\displaystyle H_2'\,K''$ & $\displaystyle - \frac{r_s \left(r - r_s\right) \left(2 r - 3 r_s\right)}{r^{4}}$ \\[1.5mm]
$\displaystyle H_2'\,H_0''$ & $\displaystyle \frac{r_s \left(r - r_s\right) \left(2 r - 3 r_s\right)}{r^{4}}$ \\[1.5mm]
$\displaystyle H_0'\,K''$ & $\displaystyle \frac{r_s \left(r - r_s\right) \left(2 r - 5 r_s\right)}{r^{4}}$ \\[1.5mm]
$\displaystyle H_0'\,H_0''$ & $\displaystyle - \frac{r_s \left(r - r_s\right) \left(2 r - 5 r_s\right)}{r^{4}}$ \\[1.5mm]
$\displaystyle K\,K''$ & $\displaystyle \frac{2 r_s \left(L - 2\right) \left(r - r_s\right)}{r^{4}}$ \\[1.5mm]
$\displaystyle K\,H_0''$ & $\displaystyle - \frac{2 r_s \left(r - r_s\right) \left(L r - 2 r + 3 r_s\right)}{r^{5}}$ \\[1.5mm]
$\displaystyle H_2\,K''$ & $\displaystyle - \frac{r_s \left(r - r_s\right) \left(L r - 4 r + 15 r_s\right)}{r^{5}}$ \\[1.5mm]
$\displaystyle H_2\,H_0''$ & $\displaystyle \frac{r_s \left(r - r_s\right) \left(L r - 4 r + 15 r_s\right)}{r^{5}}$ \\[1.5mm]
$\displaystyle H_0\,K''$ & $\displaystyle - \frac{r_s \left(r - r_s\right) \left(L r - 3 r_s\right)}{r^{5}}$ \\[1.5mm]
$\displaystyle H_0\,H_0''$ & $\displaystyle \frac{r_s \left(r - r_s\right) \left(L r + 3 r_s\right)}{r^{5}}$ \\[1.5mm]
\end{longtable}
\endgroup

\section{Exact first-order sources and scalar reduction}
\label{app:sources}

In the dimensionless system \eqref{eq:dimensionlessFirstOrderSystem}, write
\begin{equation}
 \bm s_H=\begin{pmatrix}s_{H,1}\\s_{H,2}\end{pmatrix},
 \qquad
 \bm s_{H'}=\begin{pmatrix}s_{H',1}\\s_{H',2}\end{pmatrix}.
 \label{eq:sourceVectorNotation}
\end{equation}
The components are
\begin{align}
 s_{H,1}={}&\frac{2}{x^7(x-1)^2}
 \bigg[
 6L^2x^2(x-1)^2
 +L\left(-48x^4+105x^3-54x^2-3x\right)
 \notag\\
 &\hspace{3.2cm}
 +2x^8-2x^7+x^6+72x^4-192x^3+250x^2-292x+164
 \bigg],
 \label{eq:sH1}\\
 s_{H,2}={}&-\frac{2}{x^7(x-1)}
 \bigg[
 6L^2x^2(x-1)
 +L\left(-48x^3+81x^2-30x\right)
 \notag\\
 &\hspace{3.2cm}
 +2x^7-x^6+72x^3-300x^2+454x-239
 \bigg],
 \label{eq:sH2}\\
 s_{H',1}={}&\frac{2}{x^6(x-1)}
 \bigg[
 3Lx(x-1)(2x+1)
 +x^7-x^6-12x^3+36x^2-109x+88
 \bigg],
 \label{eq:sHp1}\\
 s_{H',2}={}&-\frac{2}{x^6}
 \bigg[
 3Lx(2x-1)+x^6-12x^2+96x-97
 \bigg].
 \label{eq:sHp2}
\end{align}

To derive the scalar source, insert \eqref{eq:XKfromX0} into the second first-order equation and use the GR equation to remove $H''$. The decomposition \eqref{eq:sourceDecomposition} then has
\begin{align}
 j_H(L,x)={}&-\frac{2}{x^8(x-1)^3}
 \bigg[
 Lx(x-1)\left(x^6-120x^2+306x-184\right)
 \notag\\
 &\hspace{2.4cm}
 +2x^7+720x^3-2670x^2+3187x-1236
 \bigg],
 \label{eq:explicitjH}\\
 j_{H'}(L,x)={}&\frac{2}{x^7(x-1)^2}
 \bigg[
 -30Lx(x-1)^2
 +x^7-180x^3+1116x^2-1720x+780
 \bigg].
 \label{eq:explicitjHp}
\end{align}
These expressions follow exactly from the generic first-order system. Applying the scalar operator in \eqref{eq:scalarMasterDimensionless} to that system gives zero after \eqref{eq:explicitjH} and \eqref{eq:explicitjHp} are used. Conversely, differentiating a solution of the scalar equation, reconstructing $X_K$ with \eqref{eq:XKfromX0}, and then using \eqref{eq:X2AlgebraicDimensionless} reproduces all three metric equations.

\section{Residue algebra for the generic logarithm}
\label{app:residue}

This appendix records the algebra between the Frobenius recurrence and the compact result \eqref{eq:residueA1}. Let
\begin{equation}
 \mathfrak F(x)\equiv x(x-1)H(x)j_L[H,H'].
 \label{eq:FresidueDefinition}
\end{equation}
Before the recurrence relations are imposed, the residue at $x=0$ is
\begin{align}
 \mathop{\rm Res}_{x=0}\mathfrak F
 =4\big[{}&L a_1^2+2L a_1a_2+62L a_1a_3-109L a_1a_4
 +31L a_2^2-109L a_2a_3
 \notag\\
 &+3a_1^2+9a_1a_2-348a_1a_3+1635a_1a_4-1104a_1a_5
 \notag\\
 &-174a_2^2+1635a_2a_3-1104a_2a_4-552a_3^2\big].
 \label{eq:rawResiduePolynomial}
\end{align}
Substitution of \eqref{eq:FrobeniusCoefficients} gives
\begin{equation}
 \mathop{\rm Res}_{x=0}\mathfrak F
 =-\frac{7}{6}a_1^2L(L-2)(L-4)(L-6).
 \label{eq:residueAfterRecurrence}
\end{equation}
The $Q$-channel integrand differs from $\mathfrak F$ by the factor $2/[L(L-2)]$ from the inverse Wronskian. Hence the coefficient of $\log x$ in $X_0$ is
\begin{equation}
 \frac{2}{L(L-2)}\mathop{\rm Res}_{x=0}\mathfrak F
 =-\frac{7}{3}a_1^2(L-4)(L-6),
 \label{eq:betaBeforeNormalization}
\end{equation}
which becomes \eqref{eq:betaX0} after \eqref{eq:a1Normalization} is used.

Around the horizon, set $x=1+y$ and write
\begin{equation}
 H=b_1y+b_2y^2+b_3y^3+\cdots.
 \label{eq:horizonFrobeniusAppendix}
\end{equation}
The GR recurrence gives $b_2=(L-3)b_1/3$ and $b_3=(L^2-10L+24)b_1/24$. Inserting these coefficients into the Laurent expansion of \eqref{eq:QchannelIntegrand} removes the $y^{-1}$ term. The horizon residue is therefore zero for symbolic $L$. The remaining poles determine the regular particular data, including $X_0(1)=12b_1$.

\section{Gauge-invariance check}
\label{app:gauge}

An even-parity gauge vector is decomposed as
\begin{equation}
 \xi_\mu\dd x^\mu=\xi_aY\dd x^a+\xi Y_A\dd x^A.
 \label{eq:gaugeVector}
\end{equation}
With the conventions of \cref{eq:generalEvenParityDecomposition}, the combination $p_a$ in \eqref{eq:paDefinition} transforms as
\begin{equation}
 p_a\longrightarrow p_a-\xi_a.
 \label{eq:paGaugeTransformation}
\end{equation}
The orbit-space perturbation and angular trace transform as
\begin{equation}
 h_{ab}\longrightarrow h_{ab}-\nabla_a\xi_b-\nabla_b\xi_a,
 \qquad
 K\longrightarrow K+\frac{L}{r^2}\xi-\frac{2}{r}r^a\xi_a,
 \label{eq:metricGaugeTransformations}
\end{equation}
while $G\to G-2\xi/r^2$. Substitution in \eqref{eq:MPinvariants} gives
\begin{equation}
 \delta\widetilde h_{ab}=0,
 \qquad
 \delta\widetilde K=0,
 \label{eq:invariantsZero}
\end{equation}
and therefore $\delta\Psi_\ell^{\rm ZM}=0$.

As a direct check, set the original perturbation to zero and generate $h_{ab}$, $j_a$, $K$, and $G$ from an arbitrary pair $(\xi_a,\xi)$. The reconstructed $p_a$ equals $-\xi_a$, so both combinations in \eqref{eq:MPinvariants} vanish identically, as does $\Psi_\ell^{\rm ZM}$.

The background dressing \eqref{eq:backgroundDressing} is conceptually separate from the gauge transformation. It is required because the perturbation is defined relative to the EFT-corrected metric, while the standard Zerilli--Moncrief formula is expanded around Schwarzschild. Its contribution is fixed by the background function $b(r)$ and cannot be removed by a linear gauge choice.

\section{Quadrupolar normalization and horizon data}
\label{app:quadrupoleChecks}

The associated-Legendre convention used throughout is
\begin{equation}
 P_2^2(z)=3(1-z^2),
 \qquad
 P_2^2(2x-1)=-12x(x-1).
 \label{eq:P22normalization}
\end{equation}
The independent branch has
\begin{equation}
 Q_2^2(2x-1)=\frac{1}{5x^3}+\order(x^{-5}).
 \label{eq:Q22normalization}
\end{equation}
For the dimensionless Zerilli--Moncrief variable,
\begin{equation}
 \Psi_{P,2}^{\rm ZM}=-2x^3+\order(x^2),
 \qquad
 \Psi_{Q,2}^{\rm ZM}=\frac{1}{5x^2}+\order(x^{-3}).
 \label{eq:quadrupoleMasterBranches}
\end{equation}

Expanding the exact family \eqref{eq:quadrupoleFamily} at the horizon gives
\begin{align}
 X_0(1)&=-144,
 &
 X_2(1)&=-144,
 &
 X_K(1)&=q_0,
 \label{eq:quadrupoleHorizonValues}
\end{align}
with the first derivatives
\begin{equation}
 X_0'(1)=936-2q_0,
 \qquad
 X_K'(1)=4(q_0-144).
 \label{eq:quadrupoleFirstDerivatives}
\end{equation}
These values agree with \eqref{eq:genericHorizonData}. Equation~\eqref{eq:quadrupoleAlphaVariation} shows more directly that changing $q_0$ adds the GR growing solution and no decaying component; the no-tidal-renormalization value is therefore $q_0=-84$.

The exact master correction \eqref{eq:exactQuadrupoleZM} was obtained only after replacing the physical radial perturbation by $X_2-bH$. If one inserts $X_2$ without the background term, the coefficient of the decaying branch changes. The agreement of the resulting branch ratio with the independent spatial extraction checks that the background and perturbation expansions have been combined consistently.

\section{Exact octupolar Green-function coefficients}
\label{app:octupoleGlobal}

The horizon-anchored coefficient of the decaying branch is the elementary function
\begin{align}
 C_Q(x)={}&-140160-\frac{13680}{x^4}+\frac{80760}{x^3}
 -\frac{153480}{x^2}-\frac{49920}{x}
 +319680x
 \notag\\
 &-43200x^2+360x^3-1320x^4+1680x^5-720x^6
 -403200\log x.
 \label{eq:CQ3global}
\end{align}
The horizon-normalized coefficient of the growing branch can be written as
\begin{align}
 C_P(x)={}&182577+\frac{6}{x-1}+\frac{76}{x^6}+\frac{873}{x^5}
 -\frac{17143}{x^4}+\frac{54535}{x^3}-\frac{38285}{x^2}-\frac{204245}{x}
 \notag\\
 &+21592x-40x^2+360x^3-660x^4+360x^5
 +70080\log(x-1)-70080\log x
 \notag\\
 &+\left(360x^6-840x^5+660x^4-180x^3+21600x^2-159840x
 \right.\notag\\
 &\left.\hspace{1.8cm}+\frac{24960}{x}+\frac{76740}{x^2}
 -\frac{40380}{x^3}+\frac{6840}{x^4}\right)
 \log\!\left(\frac{x-1}{x}\right)
 \notag\\
 &+201600\log x\log\!\left(\frac{x-1}{x}\right)
 +100800(\log x)^2
 +201600\operatorname{Li}_2(1-x).
 \label{eq:CP3global}
\end{align}
The expression is real for the exterior region $x>1$. Equivalent forms involving $\operatorname{Li}_2(1/x)$ follow from standard dilogarithm identities.

The complete no-tidal-renormalization scalar correction is
\begin{equation}
 X_0^{\rm no\mbox{-}tide}
 =P_3^2(2x-1)\left[C_P(x)+c_{\rm no\mbox{-}tide}\right]
 +Q_3^2(2x-1)C_Q(x),
 \label{eq:X0globalNoTide}
\end{equation}
with $c_{\rm no\mbox{-}tide}$ given in \eqref{eq:octupoleNoTideConstant}. Differentiating \eqref{eq:X0globalNoTide} verifies the scalar equation exactly. Reconstruction of $X_K$ and $X_2$ then verifies both first-order equations and the algebraic constraint.

The separate channel product $Q_3^2C_Q$ is finite at the horizon but contains terms of the form $y^2\log y$. The corresponding logarithms in $P_3^2C_P$ cancel them. This is why horizon regularity must be imposed on the full Green-function solution rather than on each channel separately.

With the dimensionless subtraction length chosen as $x_0=1$, the finite Zerilli--Moncrief branch coefficient is
\begin{equation}
 B_3^{\rm ZM}(1)=-\frac{44431417887}{125000}\epse.
 \label{eq:octupoleFiniteZM}
\end{equation}
This number shifts under $x_0\to e^\sigma x_0$ according to \eqref{eq:ZMRGlaw}; it is quoted only to make the global solution reproducible, not as a scheme-independent observable.

\section{Asymptotic normalization and worldline checks}
\label{app:normalization}

This appendix supplies two checks used in \cref{sec:canonicalWorldline}: the leading coefficient of the growing Zerilli--Moncrief solution and the dimensional normalization of the worldline operator.

\subsection{Growing and decaying master branches}

The leading coefficient of the Legendre polynomial is
\begin{equation}
 P_\ell(z)=\frac{(2\ell)!}{2^\ell(\ell!)^2}z^\ell+\order(z^{\ell-2}).
 \label{eq:LegendreLeadingCoefficient}
\end{equation}
Using $P_\ell^2=(1-z^2)P_\ell''$, $z=2x-1$, and the GR constraint \eqref{eq:KGRconstraint}, insertion in \eqref{eq:RWZM} gives the coefficient $\mathcal A_\ell$ in \eqref{eq:ZMgrowingNormalization}. In particular,
\begin{equation}
 \Psi_{P,2}^{\rm ZM}=-2x^3+\order(x^2),
 \qquad
 \Psi_{Q,2}^{\rm ZM}=\frac{1}{5x^2}+\order(x^{-3}).
 \label{eq:quadrupoleMasterBranchesAppendix}
\end{equation}
The decaying normalization can be written as
\begin{equation}
 \mathcal N_\ell
 =\frac{(\ell+1)(\ell+2)(\ell!)^2}
 {(\ell-1)\ell(2\ell+1)!},
 \label{eq:NellFactorialForm}
\end{equation}
which is equivalent to \eqref{eq:ZMdecayingNormalization}.

As a low-multipole check, $\mathcal N_3=1/42$ and therefore
\begin{equation}
 a_3^{\log}=\mathcal N_3\beta_3^{\rm ZM}=9600\epse.
 \label{eq:ell3RawLogCheck}
\end{equation}
Equation~\eqref{eq:rawLogConversion} then gives $\beta_3^+=-15360\epse$, in agreement with \eqref{eq:canonicalOctupoleBeta}.

For $\ell=2$, both logarithmic beta functions vanish. The finite ratio is consequently not constrained by the limit
\begin{equation}
 \lim_{\ell\to2}\frac{\beta_\ell^+}{\beta_\ell^{\rm ZM}}=-\frac4{15},
 \label{eq:ell2ConversionLimit}
\end{equation}
because that limit compares logarithmic normalizations, not the finite subtraction prescription at an already integer multipole.

\subsection{Worldline convention and dimensions}

The definition \eqref{eq:worldlineOperator} implies
\begin{equation}
 c_{E,\ell}^{\rm fs}=\frac{\lambda_\ell^E}{8\pi G\ell!}.
 \label{eq:cEfromLambdaAppendix}
\end{equation}
The canonical convention for the dimensionful deformability is
\begin{equation}
 \lambda_\ell^E
 =\frac{8\pi\,2^\ell\ell!}{(2\ell)!}
 k_\ell^E R_{\rm ref}^{2\ell+1},
 \label{eq:lambdaLoveConvention}
\end{equation}
which reproduces \eqref{eq:LoveToWilson}. Since $[G]=({\rm length})^2$, $[\Mzero]={\rm length}$, and $\beta_\ell^+$ is dimensionless,
\begin{equation}
 \left[r_0\frac{\dd c_{E,\ell}^{\rm fs}}{\dd r_0}\right]
 =({\rm length})^{2\ell-1},
 \label{eq:worldlineDimensionAppendix}
\end{equation}
as required for the operator quadratic in $\mathcal E_{a_1\cdots a_\ell}$. Replacing $r_0$ by the energy scale $\mu_{\rm R}=r_0^{-1}$ reverses the sign of the renormalization-group equation but not the coefficient of the logarithm in the exterior solution.

\section{Validation hierarchy and reproducibility}
\label{app:validation}

The symbolic calculation was organized so that each reduction was checked against a logically earlier representation. This appendix summarizes the tests that support the formulas in the paper without reproducing the exploratory development history.

\begingroup
\small
\renewcommand{\arraystretch}{1.18}
\setlength{\tabcolsep}{8pt}
\setlength{\LTleft}{\fill}
\setlength{\LTright}{\fill}
\setlength{\LTcapwidth}{0.90\textwidth}
\begin{longtable}{@{}V{0.18\textwidth}V{0.54\textwidth}M{0.18\textwidth}@{}}
\caption{Algebraic checks used in the calculation. Equalities were evaluated with exact rational expressions rather than floating-point arithmetic.}
\label{tab:validationHierarchy}\\
\toprule
\multicolumn{1}{c}{\textbf{Stage}} & \multicolumn{1}{c}{\textbf{Check}} & \textbf{Result} \\
\midrule
\endfirsthead
\multicolumn{3}{c}{\itshape Table \thetable\ continued}\\[-0.15em]
\toprule
\multicolumn{1}{c}{\textbf{Stage}} & \multicolumn{1}{c}{\textbf{Check}} & \textbf{Result} \\
\midrule
\endhead
\midrule
\multicolumn{3}{r@{}}{\itshape continued on next page}\\
\endfoot
\bottomrule
\endlastfoot
Angular projection & Direct $\theta$ integration agrees with the $u=\cos\theta$ projection, including the Jacobian $1/\sqrt{1-u^2}$ at $\ell=2$ & Agreement in all three sectors \\
Generic radial action & Low-degree coefficient functions in $L$ reproduce the direct projections for $\ell=2,\ldots,13$ & All direct projections reproduced \\
GR dynamics & The variation of $\mathcal L_{\rm EH}^{(0)}(L)$ is evaluated on the GR tide and constraint & Vanishes identically in $L$ \\
Metric closure & The two first-order equations and the algebraic $X_2$ relation are substituted in the three order-$\epse$ Euler--Lagrange equations & Each equation vanishes identically \\
Scalar reduction & $X_K$ and $X_2$ reconstructed from a scalar solution are inserted in the metric system & Identity in $L$ \\
Quadrupole & The generic system reproduces the analytic family, horizon data, fixed branch ratio, and Zerilli--Moncrief correction & Closed-form agreement \\
Running proof & The Frobenius recurrence reduces the Green-function residue to \eqref{eq:residueA1} & Algebraic identity in $L$ \\
Octupole & The global $X_0$, reconstructed fields, horizon series, and asymptotic series are compared & Closed-form agreement \\
Gauge bridge & Pure-gauge perturbations give zero master variable; the Regge--Wheeler reduction agrees with the covariant formula & Algebraic agreement \\
Canonical bridge & The asymptotic branch map gives \eqref{eq:canonicalBridge}, which reduces to the canonical closed form & Symbolic agreement \\
Worldline normalization & The quadrupolar coefficient and the generic finite-size dimensions are checked against the stated convention & Agreement \\
\end{longtable}
\endgroup

\subsection{Reconstruction of the arbitrary-\texorpdfstring{$L$}{L} actions}

The projected Lagrangians have the same monomial support for all multipoles considered. Each coefficient is reconstructed as a polynomial in $L$, with degree bounded by the angular argument in \cref{sec:angularAction}. Two multipoles determine a linear coefficient and three determine a quadratic one; every remaining projection through $\ell=13$ is then a test. The symbolic field equations, scalar source, Frobenius recurrence, and residue are derived only after this closed action has been obtained. The final beta function is therefore an algebraic consequence of the reconstructed action, not an interpolation of response data.

\subsection{Ancillary files}

The ancillary files included with this submission contain the following machine-readable expressions:
\begin{description}[style=nextline,leftmargin=2.5em]
 \item[\texttt{VerifyMainResults.m}]
 Driver that loads the machine-readable expressions and checks the stored radial-system formulas, low-multipole response anchors, and normalization identities quoted in the manuscript.
 \item[\texttt{L2EH0\_GenericL.m}, \texttt{L2EH1\_GenericL.m}, \texttt{L2C3\_GenericL.m}]
 Exact radial actions in the coefficient convention of \cref{app:actions}.
 \item[\texttt{A\_GenericL.m}, \texttt{bH\_GenericL.m}, \texttt{bHp\_GenericL.m}]
 First-order matrix and source vectors before the change to $x=r/\rs$.
 \item[\texttt{X2\_Algebraic\_GenericL.m}, \texttt{J\_GenericL.m}]
 Algebraic metric reconstruction and scalar source.
 \item[\texttt{X0\_Global\_NoTide.m}]
 Exact horizon-regular octupolar scalar solution with the no-tidal-renormalization condition imposed.
 \item[\texttt{ZerilliCorrectionEll2.m}, \texttt{ZerilliCorrectionEll3.m}]
 Gauge-invariant master corrections used for the low-multipole anchors.
 \item[\texttt{BetaZM\_GenericEll.m}, \texttt{BetaCanonical\_GenericEll.m}, \texttt{BetaFiniteSizeWilson\_GenericEll.m}]
 Closed formulas for the Zerilli--Moncrief, canonical, and finite-size Wilson beta functions.
 \item[\texttt{ZMToCanonicalFactor\_GenericEll.m}]
 Multipole-dependent normalization factor relating the Zerilli--Moncrief logarithm to the canonical electric beta function.
\end{description}
The ancillary files use exact integers and rational functions throughout, so the listed checks do not require floating-point input.

\end{document}